\documentclass[aps,preprint,nofootinbib,superscriptaddress]{revtex4}%
\usepackage{amssymb}
\usepackage{amsfonts}
\usepackage{amsmath}
\usepackage{graphicx}
\usepackage[usenames]{color}
\usepackage{hyperref}%
\providecommand{\U}[1]{\protect\rule{.1in}{.1in}}
\hypersetup{
colorlinks=true,
linkcolor=blue,
citecolor=blue,
urlcolor=blue,
}

\begin{document}
\preprint{ }
\title{Extending the Feynman variational principle and analytical methods to lattice polarons}
\author{S. N. Klimin}
\affiliation{TQC, Departement Fysica, Universiteit Antwerpen, Universiteitsplein 1, B-2610
Antwerpen, Belgium}
\author{J. Tempere }
\altaffiliation[Also at ]{Lyman Laboratory of Physics, Harvard University, Cambridge, MA 02138, USA}

\affiliation{TQC, Departement Fysica, Universiteit Antwerpen, Universiteitsplein 1, B-2610
Antwerpen, Belgium}
\author{M. Houtput}
\affiliation{TQC, Departement Fysica, Universiteit Antwerpen, Universiteitsplein 1, B-2610
Antwerpen, Belgium}
\author{I. Zappacosta}
\affiliation{TQC, Departement Fysica, Universiteit Antwerpen, Universiteitsplein 1, B-2610
Antwerpen, Belgium}
\author{S. Ragni}
\altaffiliation[Also at ]{Faculty of Physics, Computational Materials Physics, University of Vienna,
Kolingasse 14-16, Vienna A-1090, Austria}

\affiliation{Department for Research of Materials under Extreme Conditions, Institute of
Physics, 10000 Zagreb, Croatia}
\author{T. Hahn}
\affiliation{Center for Computational Quantum Physics, Flatiron Institute, 162 5th Avenue,
New York, New York 10010, USA}
\author{L. Celiberti}
\affiliation{Faculty of Physics, Computational Materials Physics, University of Vienna,
Kolingasse 14-16, Vienna A-1090, Austria}
\author{C. Franchini}
\altaffiliation[Also at ]{Department of Physics and Astronomy \textquotedblleft Augusto
Righi\textquotedblright, Alma Mater Studiorum -- Universit\`{a} di Bologna,
Bologna, 40127 Italy}

\affiliation{Faculty of Physics, Computational Materials Physics, University of Vienna,
Kolingasse 14-16, Vienna A-1090, Austria}
\author{A. S. Mishchenko}
\altaffiliation[Also at ]{RIKEN Center for Emergent Matter Science (CEMS), Wako, Saitama 351-0198, Japan}

\affiliation{Department for Research of Materials under Extreme Conditions, Institute of
Physics, 10000 Zagreb, Croatia}

\begin{abstract}
We develop and systematically compare several analytical approximations for
the ground state and zero-temperature polaron dispersion in finite-width,
nonparabolic conduction bands. The main focus of the work is an extension of
the Feynman variational method to a tight-binding lattice, where the
effective-mass approximation is no longer applicable. The resulting
variational formulation is not restricted to a specific phonon dispersion or
electron-phonon interaction and provides a uniform description across weak-,
intermediate-, and strong-coupling regimes. In addition, we revisit and
generalize other analytical approaches traditionally formulated for continuum
polarons, including canonical transformations and self-consistent
Wigner-Brillouin-type approximations. For lattice polarons, these methods
exhibit qualitative features absent in the continuum case, such as a
nontrivial connection between weak- and strong-coupling limits. We show that
an improved Wigner-Brillouin scheme yields a momentum-dependent polaron
self-energy free of resonances and in good agreement with numerically exact
results over the whole range of momenta within the Brillouin zone. All methods
are applied to the Holstein model on a tight-binding lattice and are
benchmarked against numerically exact calculations, including diagrammatic
Monte Carlo (both our calculations and preceding works) and exact
diagonalization results. Furthermore, the analytical approaches are extended
to polarons with Rashba-type spin-orbit coupling, providing a stringent test
of their applicability in systems with nontrivial band structure. Our results
demonstrate that the modified Feynman variational method yields ground-state
energies and dispersions with accuracy comparable to, and in many cases
exceeding, that of other established analytical approaches. The developed
framework offers a versatile and reliable analytical description of lattice
polarons beyond the continuum approximation.

\end{abstract}
\date{\today}
\maketitle

\section{Introduction \label{Intro}}

The polaron, originally introduced by Landau as an electron self-trapped by
the polarization field of a crystal lattice \cite{Landau}, constitutes one of
the most fundamental and enduring problems in many-body physics. In its most
general form, the polaron problem describes a quantum particle interacting
with a bosonic field \cite{Froehlich,Holstein,Holstein2,Feynman}. Despite its
apparent simplicity, it has resisted an exact analytical solution and has
therefore long served as a benchmark problem for the development and testing
of analytical approximations and numerical methods \cite{FeynmanBook}.

Beyond its conceptual importance, polaron physics plays a central role in a
wide range of physical systems in which electron-phonon interactions are not
negligible. These include traditional condensed-matter settings such as ionic
crystals and molecular solids \cite{Franchini}, as well as more recent
realizations in ultracold atomic gases \cite{Shchadilova}, polymers
\cite{Ahsin}, and even dense nuclear matter \cite{Tajima}. In parallel with
these developments, the scope of polaron physics has broadened considerably,
encompassing new types of polarons, nontrivial band structures, and additional
couplings such as spin-orbit interaction \cite{Celiberti}. As a result, there
is a renewed demand for analytical approaches capable of providing both
quantitative accuracy and physical transparency across a wide range of parameters.

Over the years, numerous analytical methods have been developed for the
polaron problem. In the weak-coupling regime, these include
Rayleigh-Schr\"{o}dinger (RS) perturbation theory and self-consistent
approaches such as the Wigner-Brillouin (WB) approximation
\cite{Berciu2006,Goodvin2006,Goodvin2007,Bhattacharya}. In the strong-coupling
regime, the Lang-Firsov (LF) approximation \cite{Lang} and the adiabatic
(Pekar-type) approximation \cite{Pekar} provide controlled descriptions of
small and large polarons, respectively. Other approaches exploit variational
ansatzes, coherent phonon states \cite{Cataudella,Cataudella2},
renormalization-group techniques \cite{Jansen}, cumulant expansions
\cite{Robinson1,Robinson2}, or canonical transformations (CT) such as the
Lee-Low-Pines (LLP) method \cite{LLP} and its extensions \cite{Shchadilova}.

Among semi-analytical techniques, the momentum-average (MA) approximation
\cite{Berciu2006,Goodvin2006,Goodvin2007} stands out as one of the most
successful methods for lattice polarons with Holstein-type electron-phonon
coupling. MA is effective not only for the Holstein polaron. It was applied to
Bari\v{s}i\'{c}-Labb\'{e}-Friedel-Su-Schrieffer-Heeger (BLF-SSH) model
\cite{B1970,B1972a,B1972b,Su} in Ref. \cite{Marchand}, to the breathing model
electron-phonon coupling \cite{Lau2007} and to other momentum-dependent models
\cite{Goodv2008}. The momentum-average approximation provides remarkably
accurate results for the ground-state energy and spectral properties over a
wide parameter range and has been extensively benchmarked against numerically
exact methods. However, its accuracy deteriorates in the adiabatic regime when
the phonon frequency becomes small compared to the electronic bandwidth, and
systematic improvements are required to extend its validity toward the
large-polaron limit \cite{Goodvin2007}.

A common feature of most analytical approaches is that they are optimized
either for a parabolic conduction band of infinite width (continuum polarons)
or for specific limiting regimes of lattice models. By contrast, many
physically relevant systems are characterized by non-parabolic conduction
bands of finite width, where neither the continuum approximation nor purely
strong- or weak-coupling treatments are fully adequate. This motivates the
search for analytical methods that remain applicable across coupling regimes
and bandwidths while retaining quantitative accuracy and conceptual clarity.

The present work is devoted to the systematic development and extension of
analytical methods for polarons in finite-width, non-parabolic conduction
bands. Focusing primarily on tight-binding lattice models, we generalize
several well-known approaches originally formulated for continuum polarons to
the lattice case. Particular emphasis is placed on methods that can
interpolate between weak- and strong-coupling regimes without ad hoc assumptions.

A central result of this work is an extension of the Feynman variational
method \cite{Feynman}, originally formulated within the effective-mass
approximation, to a tight-binding conduction band of arbitrary width. This
generalized Feynman approach is not restricted to a specific form of the
electron-phonon interaction or phonon dispersion and naturally encompasses
both small- and large-polaron regimes. We show that, for lattice polarons, the
modified Feynman variational method yields ground-state energies that are at
least as accurate as those obtained from the momentum-average approximation
and, in many cases, even closer to numerically exact results.

In parallel, we revisit the method of canonical transformations for lattice
polarons. While the Lee-Low-Pines approach is traditionally regarded as a
weak-coupling method for continuum polarons, we demonstrate that its extension
to finite-width tight-binding bands leads to qualitatively new behavior.
Remarkably, the canonical-transformation approach captures not only the
weak-coupling limit but also reproduces the correct leading-order
strong-coupling (Lang-Firsov) behavior, revealing a nontrivial variational
structure absent in the continuum case.

We also reconsider the self-consistent Wigner-Brillouin approximation, which
provides a momentum-dependent polaron self-energy free of resonance
divergences. Building on earlier work by Lindemann \cite{Lindemann} and by
Gerlach and Smondyrev \cite{Gerlach}, we implement an improved WB scheme (IWB)
that exactly reproduces the RS result at zero momentum while yielding a
substantially more accurate polaron dispersion throughout the Brillouin zone.

Finally, we extend all considered methods to polarons with Rashba-type
spin-orbit coupling. Spin-orbit-coupled polarons provide a stringent test of
analytical approaches, as they combine band nonparabolicity, internal spin
structure, and electron-phonon interaction on equal footing. We show that both
the canonical-transformation method and a reduced Feynman variational approach
remain applicable in this context and yield results in good agreement with
available numerically exact calculations. For comparison, we refer here to
well-established numeric methods: the Diagrammatic Monte Carlo (DiagMC) method
applied to a polaron in both parabolic \cite{Mishchenko2000} and non-parabolic
conduction bands, including preceding works \cite{Berciu2006,Goodvin2006} and
our own calculation in the present paper; the exact diagonalization techniques
\cite{Wellein1997,Bonca1999,Li2011}.

The paper is organized as follows. In Sec.~\ref{Methods}, we present the
formulation and extension of analytical methods for polarons in finite-width
tight-binding bands, with and without Rashba spin-orbit coupling. In
Sec.~\ref{Results}, these methods are applied to lattice polarons, and their
predictions for ground-state energies and dispersions are systematically
compared with numerically exact results. Section~\ref{Conclusions} summarizes
the main findings and outlines perspectives for further developments.

\section{Analytic methods \label{Methods}}

\subsection{Electron-phonon systems \label{E-ph}}

\subsubsection{One-band polaron \label{EPh1band}}

For a polaron in a single conduction band, we consider an electron--phonon
system described by a sufficiently general Hamiltonian with linear
interaction, without restrictions on phonon frequencies and interaction
amplitudes,%
\begin{equation}
H=\varepsilon\left(  \mathbf{k}\right)  +\sum_{\mathbf{q}}\omega_{\mathbf{q}%
}\left(  b_{\mathbf{q}}^{\dag}b_{\mathbf{q}}+\frac{1}{2}\right)  +\frac
{1}{\sqrt{V}}\sum_{\mathbf{q}}V_{\mathbf{q}}e^{i\mathbf{q}\cdot\mathbf{\hat
{r}}}\left(  b_{\mathbf{q}}+b_{-\mathbf{q}}^{\dag}\right)  , \label{Hs1}%
\end{equation}
where $\omega_{\mathbf{q}}$ is the phonon frequency and $V_{\mathbf{q}}$ is
the amplitude of the electron--phonon interaction, with phonon creation and
annihilation operators $b_{-\mathbf{q}}^{\dag}$ and $b_{\mathbf{q}}$. The
Hamiltonian (\ref{Hs1}) is written in the momentum representation. We consider
the polaron on a linear, square, or cubic lattice (depending on the
dimensionality) with lattice constant $a_{0}$. Consequently, the momentum
vectors $\mathbf{k}$ and $\mathbf{q}$ belong to the reciprocal lattice. Thus,
the operator $e^{i\mathbf{q}\cdot\mathbf{\hat{r}}}$ performs a vector
displacement of a $\mathbf{k}$-dependent function according to $e^{i\mathbf{q}%
\cdot\mathbf{\hat{r}}}f\left(  \mathbf{k}\right)  =f\left(  \mathbf{k}%
-\mathbf{q}\right)  $.

The band energy in a crystal of dimensionality $D$ is chosen in the
tight-binding form%
\begin{equation}
\varepsilon\left(  \mathbf{k}\right)  =-2t\sum_{j=1}^{D}\cos k_{j}a_{0}.
\label{ek}%
\end{equation}
The parameter $t$ characterizes the conduction-band width, which is equal to
$4tD$. The system volume $V$ is related to the number of lattice cells $N$ by
$V=Na_{0}^{D}$, with $N=L^{D}$. In the wide-band limit $t\gg\omega
_{\mathbf{q}}$, the Hamiltonian (\ref{Hs1}) describes large continuum
polarons. The interaction amplitudes include various types, such as
Fr\"{o}hlich \cite{Alexandrov}, Holstein, etc. In the calculations below, we
partly focus on a more specific case, namely the Holstein interaction, with a
constant phonon frequency $\omega_{\mathbf{q}}=\omega_{0}$ and linear
interaction amplitudes given by
\begin{equation}
V_{\mathbf{q}}/\sqrt{V}=g/\sqrt{N}. \label{Holstein}%
\end{equation}
We also use the dimensionless electron--phonon coupling strength $\lambda$,
defined following Ref.~\cite{Berciu2006}:%
\begin{equation}
\lambda=\frac{1}{D}\frac{g^{2}}{\omega_{0}t}. \label{Lambda}%
\end{equation}
It should be noted that all derivations in the present work can be
straightforwardly reproduced without any change for other polaron types, as
long as they can be described by (\ref{Hs1}) with (\ref{ek}).

\subsubsection{Polaron with the spin-orbit coupling \label{EPhSO}}

In the present work, we consider a polaron in two dimensions with Rashba-type
spin--orbit coupling \cite{Rashba}. The Hamiltonian for the Rashba polaron in
two dimensions follows Ref.~\cite{Covaci2009}:
\begin{align}
H  &  =\sum_{\mathbf{k}}\left(
\begin{array}
[c]{cc}%
c_{\mathbf{k},\uparrow}^{\dag} & c_{\mathbf{k},\downarrow}^{\dag}%
\end{array}
\right)  \left(
\begin{array}
[c]{cc}%
\epsilon\left(  \mathbf{k}\right)  & \phi\left(  \mathbf{k}\right) \\
\phi^{\ast}\left(  \mathbf{k}\right)  & \epsilon\left(  \mathbf{k}\right)
\end{array}
\right)  \left(
\begin{array}
[c]{c}%
c_{\mathbf{k}\uparrow}\\
c_{\mathbf{k}\downarrow}%
\end{array}
\right) \nonumber\\
&  +\sum_{\mathbf{q}}\omega_{\mathbf{q}}b_{\mathbf{q}}^{\dag}b_{\mathbf{q}%
}+\frac{g}{\sqrt{N}}\sum_{\mathbf{k},\mathbf{q}}\left(  b_{\mathbf{q}%
}+b_{-\mathbf{q}}^{\dag}\right)  \left(  c_{\mathbf{k}+\mathbf{q},\uparrow
}^{\dag}c_{\mathbf{k}\uparrow}+c_{\mathbf{k}+\mathbf{q},\downarrow}^{\dag
}c_{\mathbf{k}\downarrow}\right)  . \label{H}%
\end{align}
The matrix elements of the free-electron part of the Hamiltonian are%
\begin{align}
\epsilon\left(  \mathbf{k}\right)   &  =-2t\left[  \cos\left(  k_{x}%
a_{0}\right)  +\cos\left(  k_{y}a_{0}\right)  \right]  ,\label{enr}\\
\phi\left(  \mathbf{k}\right)   &  =2V_{s}\left[  i\sin\left(  k_{x}%
a_{0}\right)  +\sin\left(  k_{y}a_{0}\right)  \right]  . \label{fi}%
\end{align}
Without loss of generality, we henceforth consider the case $V_{s}\geq0$.

For a single polaron, we introduce a spinor state for the electron and rewrite
(\ref{H}), according to the standard correspondence between one-particle and
second-quantized operators, in the form%
\begin{align}
H  &  =H_{e}+\left(
\begin{array}
[c]{cc}%
1 & 0\\
0 & 1
\end{array}
\right)  \left(  \sum_{\mathbf{q}}\omega_{\mathbf{q}}b_{\mathbf{q}}^{\dag
}b_{\mathbf{q}}+\frac{g}{\sqrt{N}}\sum_{\mathbf{q}}\left(  b_{\mathbf{q}%
}+b_{-\mathbf{q}}^{\dag}\right)  e^{i\mathbf{q}\cdot\mathbf{\hat{r}}}\right)
,\label{Ha}\\
H_{e}  &  =\left(
\begin{array}
[c]{cc}%
\epsilon\left(  \mathbf{k}\right)  & \phi\left(  \mathbf{k}\right) \\
\phi^{\ast}\left(  \mathbf{k}\right)  & \epsilon\left(  \mathbf{k}\right)
\end{array}
\right)  \label{Hb}%
\end{align}

The matrix Hamiltonian (\ref{Hb}) is written in the spin-up and spin-down
basis. Since a spin rotation will be used, we describe it explicitly. This
unitary transformation is performed using the matrix%
\begin{equation}
U=\frac{1}{\sqrt{2}}\left(
\begin{array}
[c]{cc}%
-\frac{\sqrt{\phi\left(  \mathbf{k}\right)  }}{\sqrt{\phi^{\ast}\left(
\mathbf{k}\right)  }} & \frac{\sqrt{\phi\left(  \mathbf{k}\right)  }}%
{\sqrt{\phi^{\ast}\left(  \mathbf{k}\right)  }}\\
1 & 1
\end{array}
\right)  . \label{U}%
\end{equation}
The transformation (\ref{U}) diagonalizes the Hamiltonian $H_{e}$ and
determines the eigenvalues and spinor eigenstates,
\begin{align}
\varepsilon_{\pm}\left(  \mathbf{k}\right)   &  =\epsilon\left(
\mathbf{k}\right)  \pm\left\vert \phi\left(  \mathbf{k}\right)  \right\vert
,\label{eigvals}\\
u_{\pm}  &  =\frac{1}{\sqrt{2}}\left(
\begin{array}
[c]{c}%
\pm\frac{\sqrt{\phi\left(  \mathbf{k}\right)  }}{\sqrt{\phi^{\ast}\left(
\mathbf{k}\right)  }}\\
1
\end{array}
\right)  . \label{eigvecs1}%
\end{align}
In contrast to the case of an electron in a single band, the band energy of an
electron with spin--orbit coupling reaches its minimum at a nonzero momentum.
For the tight-binding dispersion laws (\ref{enr}) and (\ref{fi}), the lower
spin-split energy band $\varepsilon_{-}\left(  \mathbf{k}\right)  $ contains
four equivalent minima (denoted as $\mathbf{k}_{0}$) at points in momentum
space with components%
\begin{equation}
k_{0j}=\pm\frac{1}{a_{0}}\arctan\left(  \frac{V_{s}}{\sqrt{2}t}\right)
\quad\left(  j=x,y\right)  . \label{k0j}%
\end{equation}
The value of $\varepsilon_{-}\left(  \mathbf{k}\right)  $ at these points,%
\begin{equation}
E_{0}\equiv\varepsilon_{-}\left(  \mathbf{k}_{0}\right)  =-4t\sqrt
{1+\frac{V_{s}^{2}}{2t^{2}}} \label{GS0}%
\end{equation}
is the ground-state energy of a free electron with spin--orbit coupling
\cite{Li2011}. The polaron shift of the ground-state energy is measured from
$E_{0}$.

\subsection{Extension of weak-coupling approximations \label{WCApps}}

The main result of the present work is the extension of the Feynman
variational method to a polaron in a nonparabolic conduction band of finite
width, as described below in Subsec.~\ref{Feynman}. For reference and
comparison, we also consider modifications of other analytic methods that were
previously developed for a continuum polaron. These methods are briefly
described, with attention paid to how they are altered for a polaron in a
tight-binding conduction band, both with and without spin--orbit coupling,
relative to the continuum case. As will be seen, their ranges of applicability
can also be reassessed in light of these modifications.

\subsubsection{Method of canonical transformations}

\paragraph{One-band polaron}

We formally exploit the same canonical transformations as in the LLP approach;
however, their application to a finite-width, nonparabolic band is nontrivial.
The first canonical transformation,
\begin{equation}
S_{1}=\exp\left(  -i\mathbf{Q}\cdot\mathbf{\hat{r}}\right)  ,\quad
\mathbf{Q}=\sum_{\mathbf{q}}\mathbf{q}b_{\mathbf{q}}^{\dag}b_{\mathbf{q}}
\label{S1}%
\end{equation}
leads to the Hamiltonian $\mathcal{H}=S_{1}^{-1}HS_{1}$,%
\begin{equation}
\mathcal{H}=\varepsilon\left(  \mathbf{k}-\mathbf{Q}\right)  +\sum
_{\mathbf{q}}\omega_{\mathbf{q}}\left(  b_{\mathbf{q}}^{\dag}b_{\mathbf{q}%
}+\frac{1}{2}\right)  +\frac{1}{\sqrt{V}}\sum_{\mathbf{q}}V_{\mathbf{q}%
}\left(  b_{\mathbf{q}}+b_{-\mathbf{q}}^{\dag}\right)  . \label{Htrans1}%
\end{equation}
Because $\mathcal{H}$ is coordinate-free, $\mathbf{k}\equiv\mathbf{P}$ is a
$c$-number vector associated with the total momentum of the electron--phonon
system (also referred to as the polaron momentum) $\mathbf{P}$.

For a nonparabolic band, the transformation of the electron Hamiltonian is
more complicated than in the parabolic case. However, for the tight-binding
model it can be carried out analytically, as described in
Ref.~\cite{PolaronPRB2024}. We use the known formula \cite{Fetter,Cahill}
involving a matrix $\mathbb{M}=\left\Vert M_{jj^{\prime}}\right\Vert $,
\begin{equation}
\exp\left(  \sum_{j,j^{\prime}}b_{j^{\prime}}^{\dag}\mathbb{M}_{jj^{\prime}%
}b_{j^{\prime}}\right)  =\mathtt{N}\exp\left(  \sum_{j,j^{\prime}}b_{j}^{\dag
}\left(  e^{\mathbb{M}}-1\right)  _{jj^{\prime}}b_{j^{\prime}}\right)  ,
\label{Disentang}%
\end{equation}
where $\mathtt{N}\left(  \ldots\right)  $ denotes normal ordering of creation
and annihilation operators. Applying this transformation to the polaron
Hamiltonian in a single conduction band with tight-binding dispersion
(\ref{ek}), we arrive at the expression used in our work \cite{PolaronPRB2024}%
:%
\begin{align}
\varepsilon\left(  \mathbf{P}-\mathbf{Q}\right)   &  =-t\sum_{j=1}^{D}\left[
e^{ia_{0}P_{j}}\mathtt{N}\exp\left(  \sum_{\mathbf{q}}\left(  e^{-ia_{0}q_{j}%
}-1\right)  b_{\mathbf{q}}^{\dag}b_{\mathbf{q}}\right)  \right. \nonumber\\
&  \left.  +e^{-ia_{0}P_{j}}\mathtt{N}\exp\left(  \sum_{\mathbf{q}}\left(
e^{ia_{0}q_{j}}-1\right)  b_{\mathbf{q}}^{\dag}b_{\mathbf{q}}\right)  \right]
. \label{Kinenr}%
\end{align}
This expression is still exact. Next, we apply the second canonical
transformation,%
\begin{equation}
S_{2}^{-1}=S_{2}^{\dag}=\exp\left[  -\sum_{\mathbf{q}}\left(  f_{\mathbf{q}%
}b_{\mathbf{q}}^{\dagger}-f_{\mathbf{q}}^{\ast}b_{\mathbf{q}}\right)  \right]
, \label{LLP2}%
\end{equation}
where the functions $f_{\mathbf{q}}$ are treated as variational parameters and
are chosen to minimize the energy.

The trial quantum state within the present approximation exploits the phonon
vacuum after the above canonical transformations, defined by the conditions%
\begin{equation}
b_{\mathbf{q}}\left\vert 0_{ph}\right\rangle =0,\qquad\left\langle
0_{ph}\right\vert b_{\mathbf{q}}^{\dag}=0. \label{vacLLP}%
\end{equation}
Averaging the transformed Hamiltonian with this phonon state, we obtain an
extension of the momentum-dependent polaron energy within the CT approximation
for a tight-binding conduction band,%
\[
E^{\left(  CT\right)  }\left(  \mathbf{P}\right)  =\left\langle 0_{ph}%
\left\vert S_{2}^{-1}\mathcal{H}S_{2}\right\vert 0_{ph}\right\rangle ,
\]
which is explicitly given by
\begin{align}
E^{\left(  CT\right)  }\left(  \mathbf{P}\right)   &  =-2t\operatorname{Re}%
\sum_{j=1}^{D}\exp\left(  ia_{0}P_{j}-\Phi_{j}\left(  \mathbf{P}\right)
\right) \nonumber\\
&  +\sum_{\mathbf{q}}\omega_{\mathbf{q}}f_{\mathbf{q}}^{\ast}f_{\mathbf{q}%
}+\frac{1}{\sqrt{V}}\sum_{\mathbf{q}}V_{\mathbf{q}}\left(  f_{\mathbf{q}%
}+f_{-\mathbf{q}}^{\ast}\right)  , \label{ELLP}%
\end{align}
where we introduce the function%
\begin{equation}
\Phi_{j}\left(  \mathbf{P}\right)  =\sum_{\mathbf{q}}\left(  1-e^{-ia_{0}%
q_{j}}\right)  f_{\mathbf{q}}^{\ast}\left(  \mathbf{P}\right)  f_{\mathbf{q}%
}\left(  \mathbf{P}\right)  . \label{Phij}%
\end{equation}

The optimal values of the phonon shifts $\{f_{\mathbf{q}}\}$ are obtained by
minimizing the polaron energy (\ref{ELLP}), which results in a closed, finite
set of equations for the functions $\Phi_{j}\left(  \mathbf{P}\right)  $,%
\begin{equation}
\Phi_{j}\left(  \mathbf{P}\right)  =\frac{1}{V}\sum_{\mathbf{q}}\left\vert
V_{\mathbf{q}}\right\vert ^{2}\frac{1-e^{-ia_{0}q_{j}}}{\Omega_{\mathbf{q}%
}^{2}\left(  \mathbf{P}\right)  }, \label{FjP}%
\end{equation}
with
\begin{equation}
\Omega_{\mathbf{q}}\left(  \mathbf{P}\right)  =\omega_{\mathbf{q}}%
+2t\sum_{j=1}^{D}\operatorname{Re}\left[  e^{ia_{0}P_{j}-\Phi_{j}\left(
\mathbf{P}\right)  }\left(  1-e^{-ia_{0}q_{j}}\right)  \right]  . \label{Wq}%
\end{equation}
Thus, instead of an infinite set of equations for $\{f_{\mathbf{q}}\}$, we
obtain the reduced set (\ref{FjP}) for the $D$ functions $\{\Phi_{j}\left(
\mathbf{P}\right)  \}$.

The polaron energy within the CT approach is then obtained using
$f_{\mathbf{q}}\left(  \mathbf{P}\right)  $ in the form%
\[
f_{\mathbf{q}}\left(  \mathbf{P}\right)  =-\frac{1}{\sqrt{V}}\frac
{V_{\mathbf{q}}^{\ast}}{\Omega_{\mathbf{q}}\left(  \mathbf{P}\right)  },
\]
which yields the resulting expression for the momentum-dependent polaron
self-energy:%
\begin{equation}
E^{\left(  CT\right)  }\left(  \mathbf{P}\right)  =-2t\sum_{j=1}%
^{D}\operatorname{Re}\left(  e^{ia_{0}P_{j}-\Phi_{j}\left(  \mathbf{P}\right)
}\right)  +\frac{1}{V}\sum_{\mathbf{q}}\left\vert V_{\mathbf{q}}\right\vert
^{2} \left(  \frac{\omega_{\mathbf{q}}}{\Omega_{\mathbf{q}}^{2}\left(
\mathbf{P}\right)  }-\frac{2}{\Omega_{\mathbf{q}}\left(  \mathbf{P}\right)
}\right)  . \label{ELLPres}%
\end{equation}
Here, the functions $\Phi_{j}\left(  \mathbf{P}\right)  $ are determined by
solving the $D$-dimensional set of equations (\ref{FjP}).

An analytic check of the polaron self-energy in the continuum limit,
$a_{0}\rightarrow0$, shows exact agreement with the Lee--Low--Pines result.
However, the CT approximation for a polaron with a finite-width band leads to
a different behavior in the strong-coupling regime. As follows from
(\ref{FjP}), for sufficiently strong coupling one finds $\operatorname{Re}%
\left(  \Phi_{j}\right)  \gg1$. Consequently, $\Omega_{\mathbf{q}}\left(
\mathbf{P}\right)  $ tends to $\omega_{\mathbf{q}}$, and the leading
strong-coupling contribution to the self-energy becomes%
\begin{equation}
E^{\left(  CT,SC\right)  }=-\frac{1}{V}\sum_{\mathbf{q}}\frac{\left\vert
V_{\mathbf{q}}\right\vert ^{2}}{\omega_{\mathbf{q}}}. \label{SC}%
\end{equation}
The expression (\ref{SC}) coincides with the leading term of the Lang--Firsov
strong coupling approximation. This behavior of the CT approximation is not
surprising: the Merrifield canonical transformation \cite{Merrifield}, having
some similarity with the present CT method and applied to the Holstein polaron
\cite{Cheng2008}, captures both weak- and strong coupling limits for the
ground-state energy.

In fact, the transition between the weak- and strong-coupling regimes within
the CT method is not smooth, because it occurs via a jump between two minima
of the variational functional for the self-energy at a certain critical
coupling, where the energies of these minima become equal. When approaching
the wide-band limit by gradually reducing the lattice constant $a_{0}$,
$E^{\left(  CT,SC\right)  }$ tends to $(-\infty)$, which could suggest a
divergence of the polaron energy. This divergence is, however, automatically
avoided, because the transition point for the coupling strength shifts to
arbitrarily large values as $a_{0}$ decreases. As a result, the
strong-coupling expression (\ref{SC}) does not appear in the LLP approximation
for a continuum polaron.

\paragraph{Polaron with Rashba spin-orbit coupling}

For a polaron with spin--orbit coupling, the method from \cite{PolaronPRB2024}
can be extended in the following way. We perform the same canonical
transformations as for the single-band polaron and arrive at the polaron
energy in matrix form:%
\begin{equation}
\mathbb{E}^{\left(  CT\right)  }\left(  \mathbf{P}\right)  =\left(
\begin{array}
[c]{cc}%
E_{11}^{\left(  CT\right)  }\left(  \mathbf{P}\right)  & E_{12}^{\left(
CT\right)  }\left(  \mathbf{P}\right) \\
\left(  E_{12}^{\left(  CT\right)  }\left(  \mathbf{P}\right)  \right)
^{\ast} & E_{11}^{\left(  CT\right)  }\left(  \mathbf{P}\right)
\end{array}
\right)  , \label{EpRashba}%
\end{equation}
with matrix elements%
\begin{align}
E_{11}^{\left(  CT\right)  }\left(  \mathbf{P}\right)   &
=-2t\operatorname{Re}\sum_{j=x,y}e^{ia_{0}P_{j}-\Phi_{j}\left(  \mathbf{P}%
\right)  }+\sum_{\mathbf{q}}\left(  \omega_{\mathbf{q}}f_{\mathbf{q}}^{\ast
}f_{\mathbf{q}}+\frac{g}{\sqrt{N}}\left(  f_{\mathbf{q}}+f_{\mathbf{q}}^{\ast
}\right)  \right)  ,\label{Ep11}\\
E_{12}^{\left(  CT\right)  }\left(  \mathbf{P}\right)   &  =2iV_{s}%
\operatorname{Im}e^{ia_{0}P_{x}-\Phi_{x}\left(  \mathbf{P}\right)  }%
+2V_{s}\operatorname{Im}e^{ia_{0}P_{y}-\Phi_{y}\left(  \mathbf{P}\right)  }.
\label{Ep12}%
\end{align}
In the absence of SO splitting, $V_{s}=0$, the CT polaron energy reduces to
the single-band polaron energy (\ref{ELLP}) multiplied by the identity matrix.

The polaron energy matrix is diagonalized to obtain the eigenvalues and
eigenfunctions corresponding to the lower and upper polaron branches. The
matrix (\ref{EpRashba}) has the same structure as the bare electron
Hamiltonian, and therefore the diagonalization proceeds in exactly the same
way. The energy matrix $\mathbb{E}^{\left(  CT\right)  }\left(  \mathbf{P}%
\right)  $ is thus diagonalized, yielding the spinor eigenfunctions%
\begin{equation}
u_{\pm}^{\left(  CT\right)  }\left(  \mathbf{P}\right)  =\frac{1}{\sqrt{2}%
}\left(
\begin{array}
[c]{c}%
\pm\frac{\sqrt{E_{12}^{\left(  CT\right)  }\left(  \mathbf{P}\right)  }}%
{\sqrt{\left(  E_{12}^{\left(  CT\right)  }\left(  \mathbf{P}\right)  \right)
^{\ast}}}\\
1
\end{array}
\right)  . \label{Utr}%
\end{equation}
The corresponding eigenvalues are given by expressions structurally similar to
those for the free electron,%
\begin{equation}
E_{\pm}^{\left(  CT\right)  }\left(  \mathbf{P}\right)  =E_{11}^{\left(
CT\right)  }\left(  \mathbf{P}\right)  \pm\left\vert E_{12}^{\left(
CT\right)  }\left(  \mathbf{P}\right)  \right\vert . \label{EigenCT}%
\end{equation}

The minimization of the spin-split polaron energies $E_{\pm}^{\left(
CT\right)  }\left(  \mathbf{P}\right)  $ is performed in the same way as for
the single-band polaron. As a result, we obtain the set of equations%
\begin{equation}
\Phi_{j}^{\pm}\left(  \mathbf{P}\right)  =\frac{g^{2}}{N}\sum_{\mathbf{q}%
}\frac{1-e^{-ia_{0}q_{j}}} {\left[  \Omega_{\mathbf{q}}\left(  \mathbf{P}%
\right)  \pm\delta\Omega_{\mathbf{q}}^{\left(  SO\right)  }\left(
\mathbf{P}\right)  \right]  ^{2}}, \label{Fpm}%
\end{equation}
with
\begin{equation}
\delta\Omega_{\mathbf{q}}^{\left(  SO\right)  }\left(  \mathbf{P}\right)
=-2V_{s}\frac{\sum_{j}\operatorname{Im}\left(  e^{ia_{0}P_{j}-\Phi_{j}\left(
\mathbf{P}\right)  }\right)  \operatorname{Im}\left(  e^{ia_{0}P_{j}-\Phi
_{j}\left(  \mathbf{P}\right)  } \left(  1-e^{-ia_{0}q_{j}}\right)  \right)  }
{\sqrt{\sum_{j}\left(  \operatorname{Im}e^{ia_{0}P_{j}-\Phi_{j}\left(
\mathbf{P}\right)  }\right)  ^{2}}}. \label{dWSO}%
\end{equation}
This quantity can be interpreted as the contribution of spin--orbit coupling
to the transition energy.

The optimal values of the polaron self-energies for the spin-split states are
obtained from (\ref{Ep11}) and (\ref{Ep12}) using the functions $\Phi_{j}%
^{\pm}\left(  \mathbf{P}\right)  $. The explicit expression for the polaron
self-energy then reads%
\begin{align}
E_{\pm}^{\left(  CT\right)  }\left(  \mathbf{P}\right)   &  =
-2t\operatorname{Re}\sum_{j=x,y}e^{ia_{0}P_{j}-\Phi_{j}\left(  \mathbf{P}%
\right)  }\nonumber\\
&  +\frac{g^{2}}{N}\sum_{\mathbf{q}}\left(  \frac{\omega_{\mathbf{q}}}{\left[
\Omega_{\mathbf{q}}\left(  \mathbf{P}\right)  \pm\delta\Omega_{\mathbf{q}%
}^{\left(  SO\right)  }\left(  \mathbf{P}\right)  \right]  ^{2}} -\frac
{2}{\Omega_{\mathbf{q}}\left(  \mathbf{P}\right)  \pm\delta\Omega_{\mathbf{q}%
}^{\left(  SO\right)  }\left(  \mathbf{P}\right)  }\right) \nonumber\\
&  \pm2V_{s}\sqrt{\sum_{j}\left(  \operatorname{Im}e^{ia_{0}P_{j}-\Phi
_{j}\left(  \mathbf{P}\right)  }\right)  ^{2}}. \label{Epm}%
\end{align}

According to (\ref{Fpm}), the optimization of the variational parameters
$f_{\mathbf{q}}$ is performed by minimizing the energy eigenvalues
(\ref{EigenCT}) for each eigenstate independently. In other words, each of the
two spin-split variational states is adjusted self-consistently. Physically,
this independent optimization is closely analogous to that in the well-known
variational method for the continuum Fr\"{o}hlich polaron in the
strong-coupling regime, see Refs.~\cite{Pekar,JTD}, where the polaron relaxed
excited state (RES) was thoroughly investigated. Similarly to the ground and
excited polaron states within the strong-coupling adiabatic approximation, the
spinor eigenstates $u_{+}^{\left(  CT\right)  }$ and $u_{-}^{\left(
CT\right)  }$ are not exactly orthogonal to each other. This allows for a
nonradiative relaxation of the upper spin-split polaron state. In contrast to
the well-known nonradiative relaxation of a relaxed excited state of a
single-band polaron without spin--orbit coupling \cite{JTD}, the nonradiative
transition in the present case occurs with a change of spin projection due to
the SO coupling.

\subsubsection{Improved self-consistent Wigner--Brillouin approximation
\label{Sec:IWB}}

For reference, we recall the momentum-dependent polaron self-energy within
Rayleigh--Schr\"{o}dinger (RS) perturbation theory, in which the unperturbed
states are products of the free-electron wave function $\Phi_{\mathbf{k}}$ for
an electron with momentum $\mathbf{k}$ and the free-phonon wave function%
\begin{equation}
\Psi^{\left(  N_{ph}\right)  }=\Phi_{\mathbf{k}}%
{\displaystyle\prod\limits_{\mathbf{q}}}
\left\vert n_{\mathbf{q}}\right\rangle ,
\end{equation}
where $N_{ph}$ is the number of phonons, $N_{ph}=\sum_{\mathbf{q}%
}n_{\mathbf{q}}$.

The polaron self-energy within RS perturbation theory is given by
\begin{equation}
E^{\left(  RS\right)  }\left(  \mathbf{k}\right)  =\varepsilon\left(
\mathbf{k}\right)  -\frac{1}{V}\sum_{\mathbf{q}}\frac{\left\vert
V_{\mathbf{q}}\right\vert ^{2}} {\omega_{\mathbf{q}}+\varepsilon\left(
\mathbf{k}-\mathbf{q}\right)  -\varepsilon\left(  \mathbf{k}\right)  }.
\label{RS}%
\end{equation}

The RS expression is able to reproduce the polaron dispersion only for
sufficiently small momenta, because it exhibits resonances when the energy
denominator vanishes. Another weak-coupling approximation, the self-consistent
Wigner--Brillouin (WB) approximation \cite{Gerlach}, is free of such
resonances but has other shortcomings. Within polaron theory, WB is equivalent
to the Tamm--Dankoff (TD) approach \cite{Bhattacharya} and to the
self-consistent Born approximation in the Green's-function formalism
\cite{Berciu2006,Goodvin2006,Goodvin2007}. The polaron self-energy within WB
is determined by%
\begin{equation}
E^{\left(  WB\right)  }\left(  \mathbf{k}\right)  =\varepsilon\left(
\mathbf{k}\right)  -\frac{1}{V}\sum_{\mathbf{q}}\frac{\left\vert
V_{\mathbf{q}}\right\vert ^{2}} {\omega_{\mathbf{q}}+\varepsilon\left(
\mathbf{k}-\mathbf{q}\right)  -E^{\left(  WB\right)  }\left(  \mathbf{k}%
\right)  }. \label{SCWB}%
\end{equation}

Within the TD formalism, the polaron self-energy satisfies a variational
inequality and provides an upper bound to the exact polaron self-energy $E$,%
\begin{equation}
E\leqslant E^{\left(  WB\right)  }.
\end{equation}

An advantage of WB over RS perturbation theory is that WB is free of resonance
divergences and therefore yields a qualitatively realistic polaron dispersion.
A disadvantage of WB, however, is that it predicts a higher ground-state
energy than RS. Since $E_{pol}^{\left(  WB\right)  }\left(  \mathbf{k}\right)
<\varepsilon\left(  \mathbf{k}\right)  $ and $E_{pol}^{\left(  RS\right)
}\left(  \mathbf{k}\right)  <E_{pol}^{\left(  WB\right)  }\left(
\mathbf{k}\right)  $, in particular $\left.  E_{pol}^{\left(  RS\right)
}\left(  \mathbf{k}\right)  \right\vert _{\mathbf{k}=0} <\left.
E_{pol}^{\left(  WB\right)  }\left(  \mathbf{k}\right)  \right\vert
_{\mathbf{k}=0}$. This issue was discussed in the literature, notably by
G.~Lindemann \emph{et al.}~\cite{Lindemann} and by B.~Gerlach and
M.~A.~Smondyrev \cite{Gerlach}. In Ref.~\cite{Lindemann}, an elegant procedure
was proposed to renormalize WB such that the resulting polaron self-energy at
zero momentum exactly coincides with $\left.  E_{pol}^{\left(  RS\right)
}\left(  \mathbf{k}\right)  \right\vert _{\mathbf{k}=0}$. This is achieved by
regrouping terms in the polaron Hamiltonian. For the present problem, the
reasoning of Ref.~\cite{Lindemann} can be slightly rephrased as follows. As
discussed by Epstein \cite{Epstein}, the decomposition of the total
Hamiltonian into unperturbed and perturbed parts is not unique and must be
considered separately for each specific problem. In the polaron problem, the
ground state ($\mathbf{k}=0$) is shifted by an amount $\Delta E_{0}$, which
lowers the variational ground-state energy. It turns out that the optimal
value of this shift, which yields the lowest ground-state energy, is%
\begin{equation}
\Delta E_{0}=\left.  \left(  E_{pol}^{\left(  RS\right)  }\left(
\mathbf{k}\right)  -\varepsilon\left(  \mathbf{k}\right)  \right)  \right\vert
_{\mathbf{k}=0}, \label{DE}%
\end{equation}
which coincides with the choice made in Refs.~\cite{Lindemann,Gerlach}. It is
therefore reasonable, for states with $\mathbf{k}\neq0$, to take this level as
the reference for measuring energies. This corresponds to the following
decomposition of the Hamiltonian:%
\begin{equation}
H=H_{0}^{\prime}+H_{1}^{\prime} =\left(  H_{0}+\Delta E_{0}\right)  +\left(
H_{1}-\Delta E_{0}\right)  . \label{regr}%
\end{equation}
We thus start from the unperturbed energy%
\begin{align}
H_{0}^{\prime}\Phi_{\mathbf{k}}\left\vert 0_{ph}\right\rangle  &  = \left(
\varepsilon\left(  \mathbf{k}\right)  +\Delta E_{0}\right)  \Phi_{\mathbf{k}%
}\left\vert 0_{ph}\right\rangle ,\label{unpert}\\
&  \left(  \left\vert 0_{ph}\right\rangle \equiv%
{\displaystyle\prod\limits_{\mathbf{q}}}
\left\vert 0_{\mathbf{q}}\right\rangle \right)
\end{align}
and solve the Schr\"{o}dinger equation using a Wigner--Brillouin-type
procedure. The first-order perturbation correction compensates the shift
$\Delta E_{0}$ in (\ref{unpert}). Including the second-order energy
correction, we obtain a result structurally similar to (\ref{SCWB}),%
\begin{equation}
E_{pol}^{\left(  IWB\right)  }\left(  \mathbf{k}\right)  =\varepsilon\left(
\mathbf{k}\right)  -\frac{1}{V}\sum_{\mathbf{q}}\frac{\left\vert
V_{\mathbf{q}}\right\vert ^{2}} {\omega_{\mathbf{q}}+\varepsilon\left(
\mathbf{k}-\mathbf{q}\right)  +\Delta E_{0}-E_{pol}^{\left(  IWB\right)
}\left(  \mathbf{k}\right)  }, \label{IWB}%
\end{equation}
where the denominator is analogous to that in Eq.~(15) of
Ref.~\cite{Lindemann}.

As can be seen from (\ref{IWB}) together with (\ref{DE}), the self-energy
within the improved WB approximation at $\mathbf{k}=0$ exactly reproduces the
RS perturbation result.

\subsection{Feynman variational method for a tight-binding polaron
\label{Feynman}}

\subsubsection{Polaron in a single band \label{SingleBand}}

\paragraph{Path-integral representation \label{Represent}}

In this section, we re-derive the phonon influence phase and variational
functional in the continuum path-integral representation, but with a
non-parabolic kinetic energy for the electron. In the path-integral formalism,
thermodynamic potentials (such as the free energy) are calculated via the
partition function, which in turn is written as a sum over all possible paths
$\mathbf{r}\left(  \tau\right)  $ of the electron that start and end at the
same point, weighted by the exponential factor
\begin{equation}
e^{-\beta F}=\mathcal{Z}=\int\mathcal{D}\mathbf{r}e^{-S\left[  \mathbf{r}%
\left(  \tau\right)  \right]  }.
\end{equation}
with the action functional
\begin{equation}
S\left[  \mathbf{r}\left(  \tau\right)  \right]  =\frac{1}{2}%
{\displaystyle\int\limits_{0}^{\beta}}
K_{e}\left(  \mathbf{\dot{r}}\right)  d\tau-\Phi\left[  \mathbf{r}\left(
\tau\right)  \right]  . \label{Strue}%
\end{equation}
Here $\mathbf{r}\left(  \tau\right)  $ is the electron path expressed in
imaginary time, yielding the Euclidean action, and $\beta=1/(k_{B}T)$, where
$T$ is the temperature. The kinetic energy $K_{e}\left(  \mathbf{\dot{r}%
}\right)  $ appears after integrating the phase-space path integral over the
electron momenta. For a nonparabolic conduction band with dispersion
$\varepsilon\left(  \mathbf{\hat{p}}\right)  $, the explicit analytic form of
$K_{e}\left(  \mathbf{\dot{r}}\right)  $ is not known. However, as shown
below, this knowledge is not required, because the kinetic energy does not
enter explicitly into the final expressions.

The influence phase corresponding to (\ref{Hs1}) has the well-known form
\cite{Feynman}:%
\begin{align}
\Phi\left[  \mathbf{r}\left(  \tau\right)  \right]   &  = \frac{1}{2}%
\sum_{\mathbf{q}}\frac{\left\vert V_{\mathbf{q}}\right\vert ^{2}}{V} \int%
_{0}^{\beta}d\tau\int_{0}^{\beta}d\tau^{\prime} e^{i\mathbf{q\cdot}\left(
\mathbf{r}\left(  \tau\right)  -\mathbf{r}\left(  \tau^{\prime}\right)
\right)  }\nonumber\\
&  \times\frac{\cosh\omega_{\mathbf{q}}\left(  \left\vert \tau-\tau^{\prime
}\right\vert -\frac{1}{2}\beta\right)  } {\sinh\frac{1}{2}\beta\omega
_{\mathbf{q}}}. \label{Phase}%
\end{align}

Following Feynman, the variational method proposed here considers a
\emph{partly quadratic} trial action in which the phonon degrees of freedom
are replaced by a fictitious particle with coordinate $\mathbf{r}_{f}\left(
\tau\right)  $ and quadratic kinetic energy,
\begin{equation}
S_{tr}\left[  \mathbf{r}\left(  \tau\right)  ,\mathbf{r}_{f}\left(
\tau\right)  \right]  =%
{\displaystyle\int\limits_{0}^{\beta}}
\left(  K_{e}\left(  \mathbf{\dot{r}}\right)  +\frac{m_{f}\mathbf{\dot{r}}%
_{f}^{2}}{2} +\frac{m_{f}w^{2}}{2}\left(  \mathbf{r}_{f}-\mathbf{r}\right)
^{2} \right)  d\tau, \label{Squad}%
\end{equation}
where the fictitious particle interacts harmonically with the electron.

The partition function corresponding to this model is
\begin{equation}
\mathcal{Z}_{tr}= \int\mathcal{D}\mathbf{r}\int\mathcal{D}\mathbf{r}_{f}
e^{-S_{tr}[\mathbf{r}\left(  \tau\right)  ,\mathbf{r}_{f}\left(  \tau\right)
]}.
\end{equation}
Expectation values of functionals $\mathcal{F}\left[  \mathbf{r}\left(
\tau\right)  \right]  $ of the electron path are given by
\begin{align}
\left\langle \mathcal{F}\left[  \mathbf{r}\left(  \tau\right)  \right]
\right\rangle _{tr}  &  =\frac{1}{\mathcal{Z}_{tr}} \int\mathcal{D}%
\mathbf{r}\, \mathcal{F}\left[  \mathbf{r}\left(  \tau\right)  \right]
\nonumber\\
&  \times\int\mathcal{D}\mathbf{r}_{f} e^{-S_{tr}[\mathbf{r}\left(
\tau\right)  ,\mathbf{r}_{f}\left(  \tau\right)  ]}.
\end{align}
It is evident that performing the path integral over $\mathbf{r}_{f}$ exactly
simplifies the result. The outcome of this integration still depends on the
path $\mathbf{r}\left(  \tau\right)  $ and can be expressed in terms of an
influence phase describing a quadratic retarded interaction for the electron.
The partition function of the model system becomes
\begin{align}
\mathcal{Z}_{tr}  &  =\mathcal{Z}_{f}\int\mathcal{D}\mathbf{r}\exp\left\{  -%
{\displaystyle\int\limits_{0}^{\beta}}
K_{e}\left(  \mathbf{\dot{r}}\right)  d\tau+\Phi_{\text{$0$}}\left[
\mathbf{r}\left(  \tau\right)  \right]  \right\} \nonumber\\
&  =\mathcal{Z}_{f}\int\mathcal{D}\mathbf{r} \exp\left\{  -S_{\text{$0$}%
}\left[  \mathbf{r}\left(  \tau\right)  \right]  \right\}  ,
\end{align}
with the retarded action
\begin{equation}
S_{\text{$0$}}\left[  \mathbf{r}\left(  \tau\right)  \right]  =%
{\displaystyle\int\limits_{0}^{\beta}}
K_{e}\left(  \mathbf{\dot{r}}\right)  d\tau-\Phi_{\text{$0$}}\left[
\mathbf{r}\left(  \tau\right)  \right]
\end{equation}
and the partition function of the fictitious particle
\begin{equation}
\mathcal{Z}_{f}= \int\mathcal{D}\mathbf{r}_{f} \exp\left\{  -%
{\displaystyle\int\limits_{0}^{\beta}}
\left[  \frac{m_{f}\mathbf{\dot{r}}_{f}^{2}}{2} +\frac{m_{f}w^{2}}%
{2}\mathbf{r}_{f}^{2} \right]  d\tau\right\}  . \label{FP}%
\end{equation}
Feynman restricted the trial action to a quadratic form, since only in this
case can the influence phase be calculated analytically, and obtained
\begin{align}
\Phi_{\text{$0$}}\left[  \mathbf{r}\left(  \tau\right)  \right]   &
=-\frac{m_{f}w^{3}}{4}%
{\displaystyle\int\limits_{0}^{\beta}}
d\tau%
{\displaystyle\int\limits_{0}^{\beta}}
d\tau^{\prime} \left[  \mathbf{r}\left(  \tau\right)  -\mathbf{r}\left(
\tau^{\prime}\right)  \right]  ^{2}\nonumber\\
&  \times\frac{\cosh\left[  w\left(  \left\vert \tau-\tau^{\prime}\right\vert
-\frac{\beta}{2}\right)  \right]  }{\sinh(\beta w/2)}.
\end{align}

Using Jensen's inequality
\begin{equation}
\left\langle e^{\mathcal{F}\left[  \mathbf{r}\left(  \tau\right)  \right]
}\right\rangle \geqslant e^{\left\langle \mathcal{F}\left[  \mathbf{r}\left(
\tau\right)  \right]  \right\rangle },
\end{equation}
and taking the logarithm leads to the variational inequality for the free
energy,
\begin{equation}
F\leqslant F_{tr}-F_{f}+\frac{1}{\beta}\left\langle \Phi_{\text{$0$}}\left[
\mathbf{r}\left(  \tau\right)  \right]  \right\rangle _{\text{$0$}}-\frac
{1}{\beta}\left\langle \Phi\left[  \mathbf{r}\left(  \tau\right)  \right]
\right\rangle _{\text{$0$}}. \label{Fvar}%
\end{equation}
In the zero-temperature limit, this yields the variational inequality for the
ground-state energy,
\begin{equation}
E^{\left(  GS\right)  }\leqslant E_{tr}^{\left(  0\right)  }-E_{f}^{\left(
0\right)  }+\lim_{\beta\rightarrow\infty}\frac{1}{\beta}\left\langle
\Phi_{\text{$0$}}\left[  \mathbf{r}\left(  \tau\right)  \right]  \right\rangle
_{\text{$0$}}-\lim_{\beta\rightarrow\infty}\frac{1}{\beta}\left\langle
\Phi\left[  \mathbf{r}\left(  \tau\right)  \right]  \right\rangle _{\text{$0$%
}}. \label{Evar0}%
\end{equation}

The averaged influence phase of phonons can be simplified by reducing the
number of integrations using the symmetry properties of the correlation
function \cite{Feynman}. After the standard change of variables, we arrive at
the following expression for the averaged influence phase:%
\begin{equation}
\frac{1}{\beta}\left\langle \Phi\right\rangle _{0}=\sum_{\mathbf{q}}%
\frac{\left\vert V_{\mathbf{q}}\right\vert ^{2}}{V}\int_{0}^{\beta/2}%
d\tau\left\langle e^{i\mathbf{q\cdot r}\left(  \tau\right)  }%
e^{-i\mathbf{q\cdot r}\left(  0\right)  }\right\rangle _{\mathbf{0}}%
\frac{\cosh\omega_{\mathbf{q}}\left(  \tau-\frac{1}{2}\beta\right)  }%
{\sinh\frac{1}{2}\beta\omega_{\mathbf{q}}}.
\end{equation}
The averaged influence phase of the fictitious particle is derived in the same
way and reads%
\begin{equation}
\frac{1}{\beta}\left\langle \Phi_{\text{$0$}}\left[  \mathbf{r}\left(
\tau\right)  \right]  \right\rangle _{0}=-\frac{m_{f}w^{3}}{2}%
{\displaystyle\int\limits_{0}^{\beta/2}}
d\tau\left\langle \left[  \mathbf{r}\left(  \tau\right)  -\mathbf{r}\left(
0\right)  \right]  ^{2}\right\rangle _{0}\frac{\cosh\left[  w\left(
\tau-\frac{\beta}{2}\right)  \right]  }{\sinh(\beta w/2)}. \label{infl}%
\end{equation}
In the particular case $T=0$, we obtain
\begin{equation}
\lim_{\beta\rightarrow\infty}\frac{1}{\beta}\left\langle \Phi_{\text{$0$}%
}\left[  \mathbf{r}\left(  \tau\right)  \right]  \right\rangle _{0}%
=-\frac{m_{f}w^{3}}{2}%
{\displaystyle\int\limits_{0}^{\infty}}
d\tau~e^{-w\tau}\left\langle \left[  \mathbf{r}\left(  \tau\right)
-\mathbf{r}\left(  0\right)  \right]  ^{2}\right\rangle _{0}. \label{infl0}%
\end{equation}
The averages in (\ref{infl}) and (\ref{infl0}) can be expressed in terms of
the same correlation function that appears in the averaged influence phase of
phonons:%
\begin{equation}
\left\langle \left[  \mathbf{r}\left(  \tau\right)  -\mathbf{r}\left(
0\right)  \right]  ^{2}\right\rangle _{0}=\left.  -\frac{\partial^{2}%
}{\partial\mathbf{q}^{2}}\left\langle e^{i\mathbf{q}\cdot\mathbf{r}\left(
\tau\right)  }e^{-i\mathbf{q}\cdot\mathbf{r}\left(  0\right)  }\right\rangle
_{0}\right\vert _{\mathbf{q}=0}. \label{Qav}%
\end{equation}

\paragraph{Eigenstates of the trial system \label{Eigen}}

The path-integral formalism was used above solely because it allows a shorter
and more transparent presentation of the derivations compared to the language
of ordered operators. Further on, we switch to the operator representation, in
which operator products appearing in correlation functions are ordered along
the axis of thermodynamic \textquotedblleft time\textquotedblright\ $\tau$.
The operator and path-integral representations are equivalent: the ordering of
points on the Euclidean time axis in the path-integral formalism corresponds
to time ordering in the operator formulation. Accordingly, the expressions for
the variational polaron energy have the same form in both representations
\cite{Bogolubov}.

As shown above, the variational polaron energy can be expressed in terms of
the retarded interaction that appears after exact averaging of the
electron--phonon density matrix over states of the phonon subsystem, or,
equivalently, in the path-integral formalism, after integration over phonon
paths. The averaging in the correlation functions entering the variational
functional is performed using the eigenfunctions of the trial Hamiltonian,
which in the momentum representation reads%
\begin{equation}
H_{tr}=\varepsilon\left(  \mathbf{k}\right)  +\frac{\mathbf{k}_{f}^{2}}%
{2m_{f}}+\frac{m_{f}w^{2}}{2}\left(  \mathbf{\hat{r}}-\mathbf{\hat{r}}%
_{f}\right)  ^{2}. \label{Htr0}%
\end{equation}

The averages in the variational functional for the polaron self-energy are
calculated with the eigenfunctions of the trial Hamiltonian (\ref{Htr0}). To
be consistent with the exact polaron model in a tight-binding conduction band
in $D$ dimensions described above, both $\mathbf{k}$ and $\mathbf{k}_{f}$ are
defined on a lattice with inverse lattice constant $2\pi/(a_{0}L)$. The
momentum $\mathbf{k}$ is restricted to the first Brillouin zone, $-\pi
/a_{0}\leqslant k_{j}<\pi/a_{0}$ $\left(  j=1,\ldots,D\right)  $, with
periodic boundary conditions. The momentum space associated with
$\mathbf{k}_{f}$ is infinite, in accordance with the model of a particle with
a parabolic dispersion confined by a harmonic potential.

We perform the transformation of coordinates and momenta%
\begin{equation}
\left\{
\begin{array}
[c]{c}%
\mathbf{\hat{r}}_{c}=a\mathbf{\hat{r}}+\left(  1-a\right)  \mathbf{\hat{r}%
}_{f}\\
\boldsymbol{\rho}=\mathbf{\hat{r}}-\mathbf{\hat{r}}_{f}%
\end{array}
\right.  ,\qquad\left\{
\begin{array}
[c]{c}%
\mathbf{k}_{c}=\mathbf{k}+\mathbf{k}_{f}\\
\mathbf{k}_{\boldsymbol{\rho}}=\left(  1-a\right)  \mathbf{k}-a\mathbf{k}_{f}%
\end{array}
\right.  , \label{trans}%
\end{equation}
with a real parameter $a$. This transformation is unitary for any value of
$a$. Moreover, one can always perform an additional unitary transformation
$e^{-ia\mathbf{k}_{c}\cdot\boldsymbol{\hat{\rho}}}$, such that the resulting
Hamiltonian is equivalent to that obtained by choosing $a=0$. Consequently,
the Hamiltonian (\ref{Htr0}) expressed in the new variables takes the form%
\begin{equation}
H_{tr}=\varepsilon\left(  \mathbf{k}_{\boldsymbol{\rho}}\right)
+\frac{\left(  \mathbf{k}_{\boldsymbol{\rho}}-\mathbf{k}_{c}\right)  ^{2}%
}{2m_{f}}-\frac{m_{f}w^{2}}{2}\frac{\partial^{2}}{\partial\mathbf{k}%
_{\boldsymbol{\rho}}^{2}}. \label{Htr}%
\end{equation}

The Hamiltonian (\ref{Htr}) does not contain the coordinate $\mathbf{\hat{r}%
}_{c}$. Hence, the momentum $\mathbf{k}_{c}$ is conserved and, according to
$\mathbf{k}_{c}=\mathbf{k}+\mathbf{k}_{f}$ in (\ref{trans}), coincides with
the total momentum of the trial system. This quantity can be identified with
the total polaron momentum $\mathbf{P}$ within the present approximation. The
eigenfunctions of $H_{tr}$ with $\mathbf{k}_{c}=\mathbf{P}$ are denoted
$\varphi_{\mathbf{P}}^{\left(  n\right)  }\left(  \mathbf{k}_{\boldsymbol{\rho
}}\right)  $ and satisfy
\begin{equation}
\left[  \varepsilon\left(  \mathbf{k}_{\boldsymbol{\rho}}\right)
+\frac{\left(  \mathbf{k}_{\boldsymbol{\rho}}-\mathbf{P}\right)  ^{2}}{2m_{f}%
}-\frac{m_{f}w^{2}}{2}\frac{\partial^{2}}{\partial\mathbf{k}_{\boldsymbol{\rho
}}^{2}}\right]  \varphi_{\mathbf{P}}^{\left(  n\right)  }\left(
\mathbf{k}_{\boldsymbol{\rho}}\right)  =\varepsilon_{\mathbf{P}}^{\left(
n\right)  }\varphi_{\mathbf{P}}^{\left(  n\right)  }\left(  \mathbf{k}%
_{\boldsymbol{\rho}}\right)  . \label{SchrEq}%
\end{equation}
The energy levels of the fictitious particle, which also enter the variational
functional, are obtained from (\ref{FP}) as solutions of
\begin{equation}
\left(  \frac{\mathbf{k}^{2}}{2m_{f}}-\frac{m_{f}w^{2}}{2}\frac{\partial^{2}%
}{\partial\mathbf{k}^{2}}\right)  \varphi_{f}^{\left(  n\right)  }\left(
\mathbf{k}\right)  =\varepsilon_{f}^{\left(  n\right)  }\varphi_{f}^{\left(
n\right)  }\left(  \mathbf{k}\right)  . \label{SchrF}%
\end{equation}

As can be seen from Eqs.~(\ref{SchrEq}) and (\ref{SchrF}), if $\mathbf{k}$ was
interpreted as a position variable, these equations would be equivalent to the
real-space Schr\"{o}dinger equation for a particle with an effective mass
$1/(m_{f}w^{2})$. They can be solved numerically exactly by diagonalizing the
trial Hamiltonian on a momentum-space lattice, which constitutes an advantage
of employing the momentum representation for practical calculations.

\paragraph{Variational functional \label{Varfun}}

After substitution of the trial eigenfunctions and energies, the
zero-temperature self-energy within the modified Feynman approximation takes
the form:%
\begin{equation}
E_{pol}^{\left(  F\right)  }\left(  \mathbf{P}\right)  =\varepsilon
_{\mathbf{P}}^{\left(  0\right)  }-\varepsilon_{f}^{\left(  0\right)  }%
+\lim_{\beta\rightarrow\infty}\frac{1}{\beta}\left\langle \Phi_{0}%
\right\rangle _{0}-\lim_{\beta\rightarrow\infty}\frac{1}{\beta}\left\langle
\hat{\Phi}\right\rangle _{0}, \label{Evar0a}%
\end{equation}
where the last term represents the contribution due to the electron--phonon
interaction,%
\begin{equation}
\lim_{\beta\rightarrow\infty}\frac{1}{\beta}\left\langle \hat{\Phi
}\right\rangle _{0}=\sum_{\mathbf{q}}\frac{\left\vert V_{\mathbf{q}%
}\right\vert ^{2}}{V}\sum_{n}\frac{\left\vert \sum_{\mathbf{k}}\left(
\varphi_{\mathbf{P}}^{\left(  0\right)  }\left(  \mathbf{k}\right)  \right)
^{\ast}\varphi_{\mathbf{P}+\mathbf{q}}^{\left(  n\right)  }\left(
\mathbf{k}+\mathbf{q}\right)  \right\vert ^{2}}{\omega_{\mathbf{q}%
}+\varepsilon_{\mathbf{P}+\mathbf{q}}^{\left(  n\right)  }-\varepsilon
_{\mathbf{P}}^{\left(  0\right)  }}. \label{Eph}%
\end{equation}
The averaged influence phase of the fictitious particle is obtained in an
explicit form by disentangling operator exponents in the correlation function
$\left\langle e^{i\mathbf{q}\cdot\mathbf{r}\left(  \tau\right)  }%
e^{-i\mathbf{q}\cdot\mathbf{r}\left(  0\right)  }\right\rangle _{0}$ and
subsequently expanding in a Taylor series in $\mathbf{q}$. This procedure
allows one to avoid a numerical evaluation of derivatives in momentum space in
(\ref{Qav}) in the limit $\mathbf{q}\rightarrow0$. As a result, we obtain
\begin{align}
&  \lim_{\beta\rightarrow0}\frac{1}{\beta}\left\langle \Phi_{\text{$0$}%
}\right\rangle _{0}=m_{f}w\left(  2t^{2}a_{0}^{2}\sum_{j=1}^{D}\sum_{n}%
\frac{\left\vert \left\langle \varphi_{\mathbf{P}}^{\left(  0\right)
}\left\vert \sin\left(  k_{j}a_{0}\right)  \right\vert \varphi_{\mathbf{P}%
}^{\left(  n\right)  }\right\rangle \right\vert ^{2}}{w+\varepsilon
_{\mathbf{P}}^{\left(  n\right)  }-\varepsilon_{\mathbf{P}}^{\left(  0\right)
}}\right. \nonumber\\
&  \left.  -\frac{1}{2}ta_{0}^{2}\sum_{j=1}^{D}\left\langle \varphi
_{\mathbf{P}}^{\left(  0\right)  }\left\vert \cos\left(  k_{j}a_{0}\right)
\right\vert \varphi_{\mathbf{P}}^{\left(  0\right)  }\right\rangle \right)  .
\label{Phi0}%
\end{align}

The variational parameters $m_{f}$ and $w$ for the polaron self-energy are
obtained using the variational inequality, which is substantiated for the
ground state, i.e., for zero polaron momentum:
\begin{equation}
E^{\left(  GS\right)  }\leqslant\left.  E_{pol}\left(  \mathbf{P}\right)
\right\vert _{\mathbf{P}=0}, \label{Ineq}%
\end{equation}
where $E^{\left(  GS\right)  }$ is the exact polaron ground-state energy.

\paragraph{Reduced Feynman approximation \label{RF}}

A simpler upper bound for the polaron energy can be derived without resorting
to an exact average of the trial influence phase. It is obtained by setting
the mass $m_{f}$ of the fictitious particle to $m_{f}\rightarrow\infty$.
Within this approximation, the variational polaron self-energy follows
directly from Eq. (\ref{Htr0}), in a more transparent manner than by taking
the limiting procedure in the averaged influence phase of the trial system
(\ref{Phi0}). In this limit, the trial Hamiltonian reduces to that of an
electron in a static parabolic potential, and the polaron self-energy is
obtained without any retarded action in the trial system. Thus, one arrives at
the variational functional,%
\begin{equation}
E_{pol}^{\left(  RF\right)  }=\left\langle \varphi_{0}^{\left(  0\right)
}\left\vert \varepsilon\left(  \mathbf{k}\right)  \right\vert \varphi
_{0}^{\left(  0\right)  }\right\rangle -\lim_{\beta\rightarrow\infty}\frac
{1}{\beta}\left\langle \varphi_{0}^{\left(  0\right)  }\left\vert \hat{\Phi
}\right\vert \varphi_{0}^{\left(  0\right)  }\right\rangle , \label{EvarSC1}%
\end{equation}
which results in the expression%
\begin{equation}
E_{pol}^{\left(  RF\right)  }=\left\langle \varphi_{0}^{\left(  0\right)
}\left\vert \varepsilon\left(  \mathbf{k}\right)  \right\vert \varphi
_{0}^{\left(  0\right)  }\right\rangle -\sum_{\mathbf{q}}\frac{\left\vert
V_{\mathbf{q}}\right\vert ^{2}}{V}\sum_{n}\frac{\left\vert \sum_{\mathbf{k}%
}\left(  \varphi_{0}^{\left(  0\right)  }\left(  \mathbf{k}\right)  \right)
^{\ast}\varphi_{0}^{\left(  n\right)  }\left(  \mathbf{k}+\mathbf{q}\right)
\right\vert ^{2}}{\omega_{\mathbf{q}}+\varepsilon_{0}^{\left(  n\right)
}-\varepsilon_{0}^{\left(  0\right)  }}, \label{EvarSC}%
\end{equation}
where the momentum dependence of the variational self-energy flattens to a
constant, $\mathbf{P}$-independent value. Here the trial eigenfunctions
$\varphi_{0}^{\left(  n\right)  }\left(  \mathbf{k}\right)  \equiv
\lim\limits_{m_{f}\rightarrow\infty}\varphi_{\mathbf{P}}^{\left(  n\right)
}\left(  \mathbf{k}\right)  $ and energies $\varepsilon_{0}^{\left(  n\right)
}\equiv\lim\limits_{m_{f}\rightarrow\infty}\varepsilon_{\mathbf{P}}^{\left(
n\right)  }$ are defined in the corresponding limit. This approximation can be
referred to as the reduced Feynman (RF) variational approach or, equivalently,
as an improved strong coupling approximation. The strong-coupling adiabatic
approximation is recovered from (\ref{EvarSC}) by retaining only the $n=0$
term in the sum over quantum numbers. Therefore, (\ref{EvarSC}) indeed
improves upon the strong coupling adiabatic ansatz and yields a lower
variational polaron energy. However, the variational functional within the
reduced Feynman approximation, $E_{pol}^{\left(  RF\right)  }$, is unable to
describe the momentum dependence of the polaron energy and cannot lie below
$E_{var}$ given by (\ref{Evar0}), because the limit $m_{f}\rightarrow\infty$
is, in general, not an optimal choice.

Despite the fact that the full variational expression for the polaron
self-energy is available, the reduced Feynman approximation is also of
interest, especially for polarons with spin--orbit coupling, because its
extension to the spin--orbit coupled case is rigorously justified, whereas the
full Feynman variational method requires a further analysis of the validity of
the variational principle for a complex Hamiltonian. Additionally, it is much
less time-consuming to evaluate in practice. Indeed, the trial eigenfunctions
$\varphi_{\mathbf{P}}^{(n)}(\mathbf{k})$ must only be evaluated at a single
momentum $\mathbf{P}$, which requires only one solution of the Sch\"{r}%
odinger-like equation in \eqref{SchrEq}. This is in contrast to the full
Feynman variational method, which requires solving this equation for every
value of $\mathbf{P}$.

Remarkably, the reduced Feynman approximation exactly reproduces both the
strong-coupling and weak-coupling limits of (\ref{Evar0}) for the ground-state
energy, and predicts only slightly higher values than the full Feynman
approach in the intermediate coupling regime. The successful application of
the reduced Feynman approximation in the weak-coupling regime may appear
counterintuitive at first sight, but it follows from the completeness of the
set of eigenstates $\left\{  \varphi_{0}^{\left(  n\right)  }\left(
\mathbf{k}\right)  \right\}  $.

\subsubsection{Polaron with spin-orbit coupling \label{RFSO}}

In the present work, we apply the reduced Feynman variational approach to the
spin--orbit coupled polaron in a tight-binding conduction band, as described
by (\ref{Ha}) and (\ref{Hb}) together with (\ref{enr}) and (\ref{fi}). The
full Feynman approach is also applicable to this problem, but it requires a
rigorous justification of the variational principle, because the
electron--phonon system is described by a complex matrix Hamiltonian which
might invalidate the Jensen inequality. In contrast, the reduced Feynman
approach is in fact based on the Ritz variational principle, which is valid
for any Hermitian Hamiltonian. Indeed, the limiting transition $m_{f}%
\rightarrow+\infty$ reduces the trial Hamiltonian (\ref{Htr}) to one with a
static quadratic interaction. As a consequence, no retarded interaction with
the fictitious particle appears in the variational functional. With respect to
the averaged retarded interaction with phonons, the same term as in
Ref.~\cite{Brosens1992} is employed, where it is shown that the Ritz
variational principle can be applied to a system describing an electron with a
retarded self-interaction generated by the electron--phonon interaction.

For a polaron with spin--orbit coupling, we employ the following trial
Hamiltonian:%
\begin{equation}
H_{tr}=\left(
\begin{array}
[c]{cc}%
\epsilon\left(  \mathbf{k}\right)  & \phi\left(  \mathbf{k}\right) \\
\phi^{\ast}\left(  \mathbf{k}\right)  & \epsilon\left(  \mathbf{k}\right)
\end{array}
\right)  -\frac{\varpi^{2}}{2}\left(
\begin{array}
[c]{cc}%
1 & 0\\
0 & 1
\end{array}
\right)  \frac{\partial^{2}}{\partial\mathbf{k}^{2}}, \label{HtrSp}%
\end{equation}
with variational parameter $\varpi$. The eigenvalues $\varepsilon_{n}$ and the
two-dimensional spinor eigenfunctions $u_{n}\left(  \mathbf{k}\right)  $ are
determined from%
\begin{equation}
H_{tr}u_{n}\left(  \mathbf{k}\right)  =\varepsilon_{n}u_{n}\left(
\mathbf{k}\right)  . \label{Eigensp}%
\end{equation}
This equation is solved numerically exactly in the same manner as above, by
exact diagonalization of the Hamiltonian (\ref{HtrSp}) on a momentum-space
lattice. The only difference with respect to the reduced Feynman approach for
a single-band polaron is the spinor, rather than scalar, form of the trial
Hamiltonian. Consequently, the resulting variational functional for the
polaron self-energy $E_{var}^{\left(  SO\right)  }$ is given by an expression
analogous to (\ref{EvarSC}), using the obtained energy eigenvalues and spinor
eigenfunctions:%
\begin{align}
E_{pol}^{\left(  RF\right)  }  &  =\sum_{\mathbf{k}}u_{0}^{\dag}\left(
\mathbf{k}\right)  \cdot\left(
\begin{array}
[c]{cc}%
\epsilon\left(  \mathbf{k}\right)  & \phi\left(  \mathbf{k}\right) \\
\phi^{\ast}\left(  \mathbf{k}\right)  & \epsilon\left(  \mathbf{k}\right)
\end{array}
\right)  \cdot u_{0}\left(  \mathbf{k}\right) \nonumber\\
&  -\sum_{\mathbf{q}}\frac{\left\vert V_{\mathbf{q}}\right\vert ^{2}}{V}%
\sum_{n}\frac{\left\vert \sum_{\mathbf{k}}u_{0}^{\dag}\left(  \mathbf{k}%
\right)  \cdot u_{n}\left(  \mathbf{k}+\mathbf{q}\right)  \right\vert ^{2}%
}{\omega_{\mathbf{q}}+\varepsilon_{n}-\varepsilon_{0}}. \label{EvarSO}%
\end{align}

For the subsequent numerical calculations, we employ nondispersive optical
phonons with frequency $\omega_{\mathbf{q}}=\omega_{0}$ and the Holstein
electron--phonon interaction described by (\ref{Holstein}) and (\ref{Lambda}).

\section{Results and discussion \label{Results}}

\subsection{Polaron ground state energy \label{NumGS}}

\subsubsection{Single band \label{NumSingle}}

\begin{figure}[ptbh]%
\centering
\includegraphics[
height=3.8467in,
width=5.175in
]%
{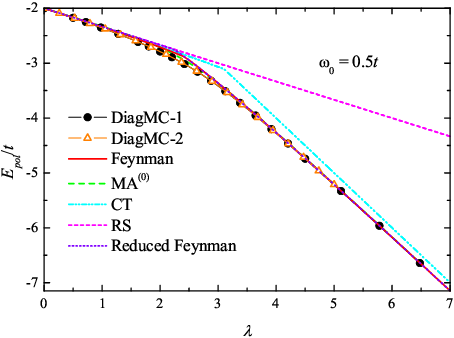}%
\caption{Ground state energy of a Holstein polaron in 1D for $\omega_{0}%
=0.5t$, calculated by different methods: Diagrammatic Monte Carlo
\cite{Berciu2006,Goodvin2006} (\emph{full dots}), Diagrammatic Monte Carlo of
the present work (\emph{triangles}), the modified Feynman variational method
(\emph{solid curve}), the reduced Feynman variational method (\emph{dotted
curve}), the Momentum Average approximation MA$^{\left(  0\right)  }$
(\emph{dashed curve}) \cite{Berciu2006,Goodvin2006}, method of canonical
transformations (\emph{dot-dot-dashed curve}), and the RS perturbation theory
\cite{Berciu2006,Goodvin2006} (\emph{short-dashed line}).}%
\label{fig:GSEn1}%
\end{figure}

In Fig.~\ref{fig:GSEn1}, we plot the ground-state energy of a Holstein polaron
in 1D as a function of the coupling constant $\lambda$ for $\omega_{0}=0.5t$,
calculated using the modified Feynman variational method described above, both
full and reduced versions. The result is compared with
Rayleigh--Schr\"{o}dinger (RS) perturbation theory, our Diagrammatic Monte
Carlo results and those of Refs. \cite{Berciu2006,Goodvin2006}, the
momentum-average approximation \cite{Berciu2006,Goodvin2006,Goodvin2007}, and
the method of canonical transformations.%

In the numerical evaluation of the polaron self-energy using the full modified
Feynman variational method with the averaged influence phase (\ref{Phi0}), one
technical point arises. This issue follows from (\ref{infl0}), which is
asymptotically exact in the limit $N\rightarrow\infty$. In the weak-coupling
limit at $N\rightarrow\infty$, the variational ground-state energy
analytically tends to the result of RS perturbation theory, analogously to the
weak-coupling regime of the continuum polaron within the Feynman variational
method \cite{Feynman}. For finite $N$, the resulting self-energy is
numerically reliable as long as the radius of the polaron ground state remains
much smaller than the system size.

In the weak-coupling limit the polaron extends over the entire crystal. As
$\lambda\rightarrow0$, the above condition ceases to hold, and the variational
functional loses stability when $\lambda$ becomes sufficiently small.
Consequently, the polaron self-energy must be evaluated starting from a finite
value of the coupling constant that ensures fulfillment of the variational
inequality. In practice, this defines a practical lower bound on the coupling
strength determined by the onset of numerical instability.

\begin{figure}[ptbh]%
\centering
\includegraphics[
height=3.845in,
width=5.3618in
]%
{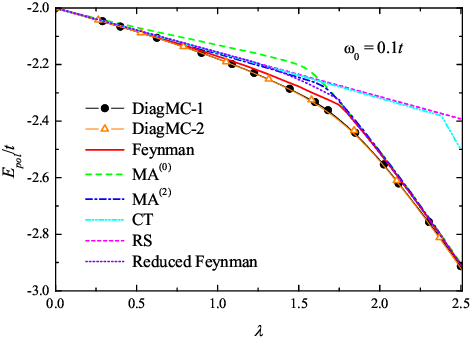}%
\caption{Ground state energy of a Holstein polaron in 1D for $\omega_{0}%
=0.1t$, calculated by different methods: Diagrammatic Monte Carlo
\cite{Berciu2006,Goodvin2006} (\emph{full dots}), Diagrammatic Monte Carlo of
the present work (\emph{triangles}), the modified Feynman variational method
(\emph{solid curve}), the reduced Feynman variational method (\emph{dotted
curve}), the momentum-average approximation MA$^{\left(  0\right)  }$ and
MA$^{\left(  2\right)  }$ (\emph{dashed and dot-dashed curves, respectively})
\cite{Berciu2006,Goodvin2006}, method of canonical transformations
(\emph{dot-dot-dashed curve}), and the RS perturbation theory
\cite{Berciu2006,Goodvin2006} (\emph{short-dashed line}).}%
\label{fig:GSEn2}%
\end{figure}

Increasing the number of lattice sites shifts this threshold to smaller
values, ensuring convergence of the lattice polaron model toward the continuum
limit. At the same time, the computational time scales as $L^{2}$. In the
present calculations we use $L=40$, for which the full-Feynman variational
functional exhibits a stable minimum down to $\lambda\approx0.6$ for
$\omega_{0}=0.5$ and down to $\lambda\approx0.5$ for $\omega_{0}=0.1$. This
practical lower bound on the coupling strength is not essential. As can be
seen from Fig.~\ref{fig:GSEn1} and, below, from Fig.~\ref{fig:GSEn2}, the
full-Feynman ground-state energy at these coupling strengths is already
extremely close to the weak-coupling perturbation result. Moreover, the number
of lattice sites can always be increased further, the only limitation being
the computational cost.

As can be seen from Fig.~\ref{fig:GSEn1}, the full version of the modified
Feynman variational method predicts the polaron ground-state energy very close
to the Diagrammatic Monte Carlo results. In the strong-coupling regime, the
variational ground-state energy obtained within the modified Feynman method is
not asymptotically exact because of the parabolic form of the trial potential
(as is also the case for the continuum polaron; see
Refs.~\cite{Miyake,Mishchenko2000}). Nevertheless, it yields ground-state
energies extremely close to the exact numerical values obtained from
Diagrammatic Monte Carlo calculations.

The reduced Feynman variational method, despite its simplicity, yields polaron
ground-state energies that are only slightly higher than those obtained from
the full formulation in the intermediate-coupling regime, spanning the
crossover between the weak- and strong-coupling limits. The results of the
reduced Feynman method are comparable to those of the full Feynman variational
approach and diagrammatic Monte Carlo.

The momentum-average approximation \cite{Berciu2006,Goodvin2006,Goodvin2007}
performs reasonably well for the parameters chosen in Fig.~\ref{fig:GSEn1}.
The variational results of the present work are of the same accuracy as the
ground-state energy predicted by the momentum-average approximation. Moreover,
in the intermediate and weak coupling range, $\lambda\lesssim1.5$, the MA
ground-state energy lies slightly above that predicted by perturbation theory,
indicating a certain weakness of MA in this regime. In contrast, the modified
Feynman method yields energies systematically lower than those obtained from
perturbation theory and the method of canonical transformations.%

Figure~\ref{fig:GSEn2} shows the polaron ground-state energy for a lower
phonon frequency, $\omega_{0}=0.1t$, for which Diagrammatic Monte Carlo data
are also available \cite{Goodvin2006,Goodvin2007} for comparison, together
with the results of different versions of the momentum-average approximation.
In particular, the improved MA$^{\left(  2\right)  }$ scheme
\cite{Goodvin2007} is included, which was designed to extend the applicability
of MA to lower phonon frequencies.

As can be seen from Fig.~\ref{fig:GSEn2}, the modified Feynman variational
method becomes more accurate than both MA$^{\left(  0\right)  }$ and
MA$^{\left(  2\right)  }$ at lower phonon frequencies. MA was considered to be
unsuitable at low phonon frequencies also for the double-well potential
\cite{Adolphs2014,Ragni1}. The poor performance of MA in the intermediate and
weak-coupling regime at $\omega_{0}=0.1t$ is more pronounced than for
$\omega_{0}=0.5t$, whereas the modified Feynman method remains in good
agreement with Diagrammatic Monte Carlo results in both figures. This behavior
can be transparently understood from the fact that the modified Feynman
variational method is not restricted to phonon frequencies that are not small
compared to $t$, and explicitly reduces to the well-known Feynman approach for
a continuum polaron in the adiabatic limit $t\gg\omega_{0}$. In addition, it
is worth noting that both the full Feynman approximation and the reduced one
provide lower polaron energies with respect to MA$^{\left(  0\right)  }$ and
MA$^{\left(  2\right)  }$ for $\omega_{0}=0.1t$.%

\begin{figure}[ptbh]%
\centering
\includegraphics[
height=5.4604in,
width=4.2471in
]%
{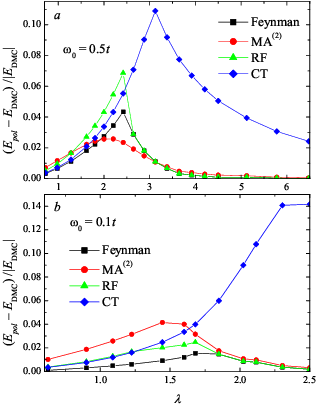}%
\caption{Relative difference between the polaron ground state energy $E_{pol}$
calculated using different analytic methods and the DMC data, with the same
parameters as in Figs. \ref{fig:GSEn1} and \ref{fig:GSEn2}. Squares: the full
Feynman approximation; circles: the Momentum Average approximation
MA$^{\left(  2\right)  }$; triangles: the reduced Feynman approximation;
diamonds: the method of canonical transformations.}%
\label{fig:DiffEnrs}%
\end{figure}

The differences between the results obtained within different approximations,
as shown in Figs.~\ref{fig:GSEn1} and \ref{fig:GSEn2}, are rather small and in
some cases hardly resolvable by eye. Consequently, we include
Fig.~\ref{fig:DiffEnrs}, where the relative difference $\left(  E_{pol}%
-E_{DMC}\right)  /\left\vert E_{DMC}\right\vert $ between the polaron energy
$E_{pol}$ and the DMC data $E_{DMC}$ is plotted as a function of the coupling
constant $\lambda$ for several analytic approximations: the modified Feynman
variational method (full and reduced versions), the improved Momentum Average
approximation MA$^{\left(  2\right)  }$, and the method of canonical
transformations. The variational inequality $E_{pol}\geqslant E_{pol}^{\left(
exact\right)  }$ (where $E_{pol}^{\left(  exact\right)  }$ is the exact
polaron ground-state energy) should hold for all the aforesaid analytic
approximations. As can be seen from the figure, $E_{pol}>E_{DMC}$ everywhere,
which numerically confirms the variational inequality for these approximations
and, most importantly for us, numerically verifies the validity of the
variational principle for the modified Feynman approximation.

As the coupling strength increases continuously, the aforesaid relative
difference for the modified Feynman method exhibits a sharp transition at a
certain value of $\lambda$, denoted by $\lambda_{c}$, to a regime in which the
full and reduced versions of the Feynman approximation coincide for stronger
couplings. This behavior is a feature of the approximation and bears some
similarity to the switch between the weak-coupling and strong-coupling regimes
in the CT approximation. However, the ground-state energy within the modified
Feynman method for $\lambda<\lambda_{c}$ is lower than the CT or perturbative
result and depends on $\lambda$ nonlinearly, so that it does not coincide with
the weak-coupling approximation below $\lambda_{c}$. In addition, the Feynman
polaron ground-state energy for $\lambda>\lambda_{c}$ differs from the
leading-order Lang-Firsov strong-coupling result, being extremely close to the
next-to-leading-order Lang-Firsov expression given below in Eq.~(\ref{LF}).
Thus, the switch between the two regimes at $\lambda_{c}$ in the Feynman
method is less abrupt than in the CT method.

For an intermediate value of the adiabatic ratio $\omega_{0}/t=0.5$, as shown
in Fig.~\ref{fig:DiffEnrs}~(a), the overall accuracy of the full modified
Feynman approximation over the entire interval of coupling strengths is close
to that of the improved Momentum Average approximation. In the adiabatic
regime with $\omega_{0}/t=0.1$, both MA$^{\left(  2\right)  }$ and the
modified Feynman method exhibit nearly the same accuracy, with the modified
Feynman method being only slightly closer to the DMC data. In the
weak-coupling regime, the modified Feynman method in the adiabatic limit
provides better agreement with DMC than MA$^{\left(  2\right)  }$.

As $\lambda$ decreases below $\lambda_{c}$, the results of the full and
reduced modified Feynman approximations tend toward each other, as well as
toward the CT and perturbative results, becoming asymptotically identical in
the weak-coupling limit. For the regimes considered in Fig.~\ref{fig:DiffEnrs}%
, the switching coupling strength $\lambda_{c}$ is not very large, and,
consequently, the convergence of the different regimes toward each other is
rather fast. This may explain why the reduced Feynman approximation performs
so well. Nevertheless, Fig.~\ref{fig:DiffEnrs} shows that for $\lambda$ close
to $\lambda_{c}$, the difference in accuracy between the full and reduced
Feynman approximations is not negligible.%

\begin{figure}[ptbh]%
\centering
\includegraphics[
height=6.5155in,
width=3.8623in
]%
{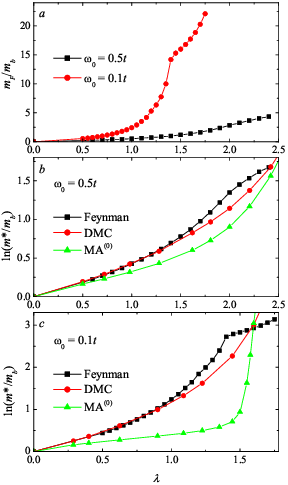}%
\caption{(\emph{a}) The parameter $m^{\ast}/m_{b}=1+m_{f}/m_{b}$ as a function
of the coupling strength $\lambda$ for $\omega_{0}=0.5t$ (squares) and
$\omega_{0}=0.1t$ (circles). The other parameters are the same as in figures
above. (\emph{b}, \emph{c}) The ratio $m^{\ast}/m_{b}$ on a logarithmic scale,
compared with the DMC and MA$^{\left(  0\right)  }$ results from
Ref.~\cite{Goodvin2006}.}%
\label{fig:Pmass}%
\end{figure}

In addition, we analyze the dependence of the mass $m_{f}$ of the fictitious
particle introduced in (\ref{Squad}) on the coupling strength for
$\lambda<\lambda_{c}$, where the optimal value of $m_{f}$ is finite. In
Fig.~\ref{fig:Pmass}~(a), we plot the parameter $m^{\ast}/m_{b}=1+m_{f}/m_{b}%
$, where $m_{b}$ is the effective mass of the electron near the bottom of the
conduction band. For a tight-binding band, $m_{b}=1/\left(  2ta_{0}%
^{2}\right)  $. The parameter $m^{\ast}$ has the meaning of the total mass of
the trial system. Upon approaching $\lambda_{c}$, $m^{\ast}/m_{b}$ becomes
rather large, which results in a relatively small difference between the
polaron self-energies obtained within the full and reduced Feynman
approximations. This further supports the explanation given above as to why
the reduced Feynman approximation appears to perform rather well despite its simplicity.

As found by Feynman \cite{Feynman}, $m^{\ast}=m_{b}+m_{f}$ differs only by a
few percent from the polaron mass derived using an expansion in powers of an
imaginary velocity in the path-integral formalism. Here, we use $m^{\ast}$ as
an estimate of the polaron mass, because it is not yet clear whether Feynman's
trick with the imaginary velocity can be reproduced for a polaron in a
nonparabolic band. In Fig.~\ref{fig:Pmass}~(b, c), $m^{\ast}/m_{b}$ is plotted
on a logarithmic scale and compared with the DMC and MA$^{\left(  0\right)  }$
numerical data from Ref.~\cite{Goodvin2006}. When the coupling strength is not
very large ($\lambda\lessapprox1.5$ for $\omega_{0}=0.5t$ and $\lambda
\lessapprox1.0$ for $\omega_{0}=0.1t$), the parameter $m^{\ast}$ is close to
the DMC result, even closer than MA$^{\left(  0\right)  }$. Of course,
improvements of MA should provide better agreement with DMC than MA$^{\left(
0\right)  }$, but we do not yet possess such results for the polaron mass.%

\begin{table}[h] \centering
\begin{tabular}
[c]{|l|l|l|l|l|l|}\hline
$n$ & $L_{n}$ & $E_{pol}/t$ & $m_{f}/m_{b}$ & $w/t$ & $\left\vert
\frac{E_{pol}\left(  L_{n}\right)  }{E_{pol}\left(  L_{n-1}\right)
}-1\right\vert $\\\hline\hline
$1$ & $20$ & $-2.71518$ & $1.36845$ & $0.750464$ & $-$\\\hline
$2$ & $30$ & $-2.71141$ & $1.40863$ & $0.737842$ & $1.388\times10^{-3}%
$\\\hline
$3$ & $40$ & $-2.71007$ & $1.42683$ & $0.732849$ & $4.942\times10^{-4}%
$\\\hline
$4$ & $60$ & $-2.70909$ & $1.44186$ & $0.72909$ & $3.616\times10^{-4}$\\\hline
$5$ & $80$ & $-2.70875$ & $1.44762$ & $0.727737$ & $1.255\times10^{-4}%
$\\\hline
\end{tabular}
\caption{Polaron ground sate energy and optimal values of the variational parameters
$m_{f}$ and $w$ within the modified Feynman variational method for $\omega
_{0}=0.5t$, and $\lambda=2.0$.}\label{Tab1}%
\end{table}%

In order to examine the convergence of the modified Feynman approximation with
an increasing number of lattice sites, we explore the evolution of the optimal
values of variational parameters for solutions of Eqs.~(\ref{SchrEq}) and
(\ref{SchrF}), as well as of the resulting ground-state energy, as the number
of sites increases. The calculation is performed for $P=0$, $\omega_{0}=0.5t$,
and $\lambda=2.0$. The results, labeled by the index $n$, are listed in
Table~\ref{Tab1}. As can be seen from the table, the relative difference
between two subsequent values of the ground-state energy decreases with
increasing number of sites, as expected.

\subsubsection{Polaron with spin-orbit coupling \label{NumSO}}

For the ground-state energy of a single-band polaron, the full Feynman
approximation exhibits numerically only a relatively small lowering of the
polaron energy in the crossover region between the weak- and strong-coupling
regimes compared to the reduced Feynman approach. In this respect, the reduced
Feynman approximation is particularly relevant for the polaron with
spin--orbit coupling, because the validity of the full Feynman approach for a
polaron described by a matrix Hamiltonian with complex matrix elements still
requires careful examination.%

\begin{figure}[ptbh]%
\centering
\includegraphics[
height=5.521in,
width=3.9228in
]%
{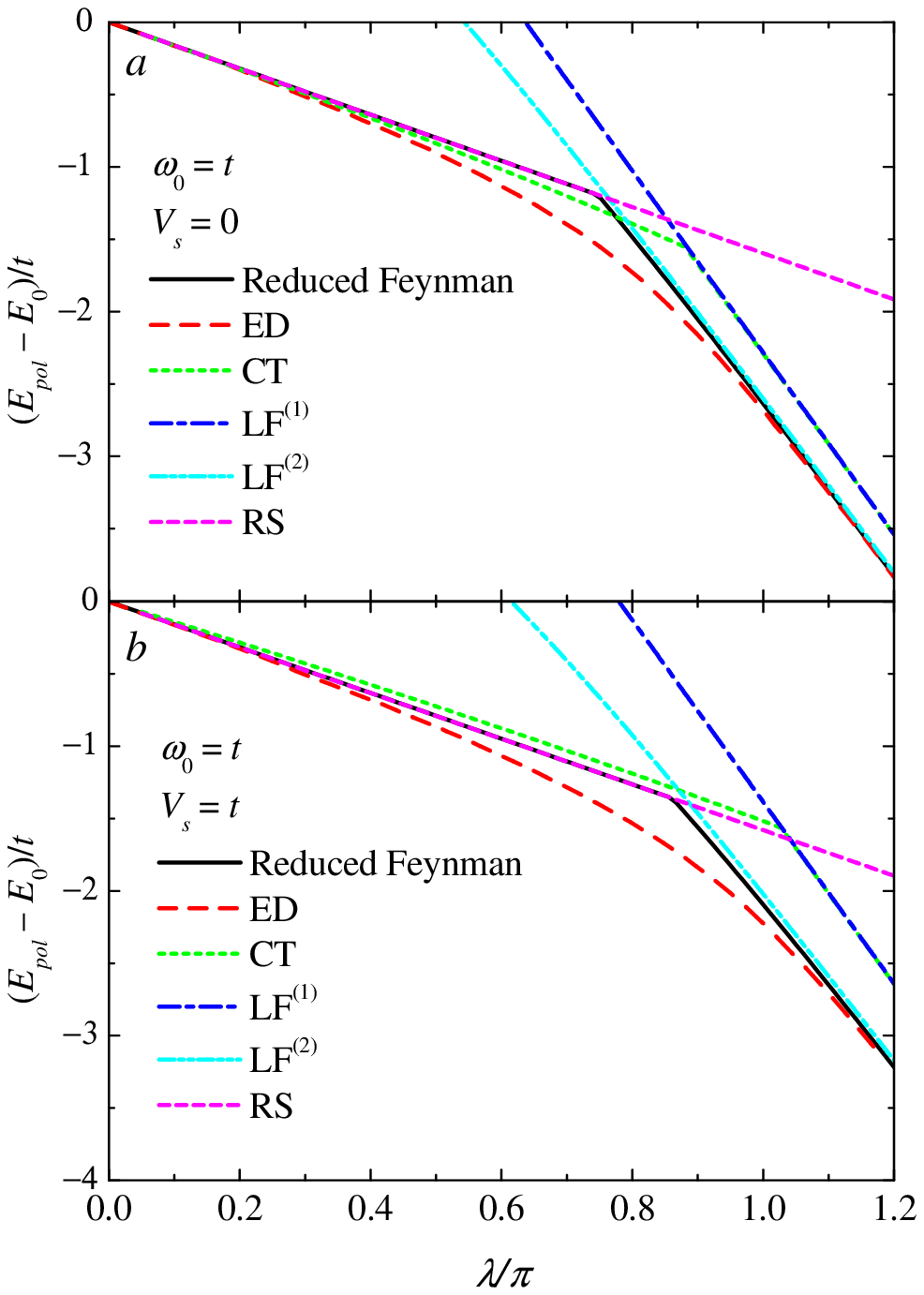}%
\caption{Ground state energy of a polaron with the Rashba spin-orbit coupling
with the phonon frequency $\omega_{0}=t$ and the SO potential $V_{s}=0$
(\emph{a}) and $V_{s}=t$ (\emph{b}) calculated using different methods: the
reduced Feynman approach (\emph{solid curves}), the exact diagonalization
method \cite{Li2011} (\emph{dashed curves}), the method of canonical
transformations (\emph{dotted curves}), the RS perturbation theory
(\emph{dot-dashed curves}), the next-to-leading order (\emph{dot--dot-dashed
curves}) and leading-order (\emph{short-dashed curves}) Lang-Firsov
approximations.}%
\label{GSSO-1}%
\end{figure}

Figure~\ref{GSSO-1} shows the difference between the ground-state energy of a
polaron with Rashba spin--orbit coupling and the free-electron ground-state
energy $E_{0}$ given by (\ref{GS0}), using the parameters of
Ref.~\cite{Li2011}: phonon frequency $\omega_{0}=t$ and spin--orbit coupling
potential $V_{s}=0$ (\emph{a}) and $V_{s}=t$ (\emph{b}). The results are
obtained using several approaches: the reduced Feynman variational method, the
numerical calculation of Ref.~\cite{Li2011} based on exact diagonalization
(ED) of the electron--phonon Hamiltonian, the method of canonical
transformations developed in the present work, Rayleigh--Schr\"{o}dinger
perturbation theory, and two variants of the Lang--Firsov approximation. The
latter include the next-to-leading-order LF approximation described in
Ref.~\cite{Li2011},
\begin{equation}
E_{LF}^{\left(  2\right)  }-E_{0}=-2t\lambda\left(  1+2\frac{t^{2}+V_{s}^{2}%
}{\left(  2t\lambda\right)  ^{2}}\right)  , \label{LF}%
\end{equation}
and the leading-order term of the LF approximation, $(-2t\lambda)$. The
coupling constant $\lambda$ in the figure is rescaled by a factor $1/\pi$ to
allow a transparent comparison, so that $\lambda$ is defined consistently in
the present work and in Ref.~\cite{Li2011}.

As can be seen from Fig.~\ref{GSSO-1}, the validity of the variational
inequality for the functional derived within the reduced Feynman approximation
is numerically confirmed for the polaron with spin--orbit coupling. The
variational polaron ground-state energy exhibits a continuous but non-smooth
crossover between the weak- and strong-coupling regimes. In fact, the
numerical calculation of Ref.~\cite{Li2011} shows no polaron \textquotedblleft
phase transition\textquotedblright\ as the coupling constant is varied.
Nevertheless, this artifact of the reduced Feynman variational method can be
useful, as it clearly highlights the coupling strength range in which the
polaron crosses over from the weak- to the strong-coupling regime.

The other analytic approximations shown in Fig.~\ref{GSSO-1} yield results
close to those obtained from the reduced Feynman approximation and from the
numerical calculation in their respective regimes of validity. RS perturbation
theory is applicable only in the weak-coupling regime. The Lang--Firsov
approximation LF$^{\left(  2\right)  }$ performs well in the strong-coupling
regime. The method of canonical transformations captures the entire range of
coupling strengths, being asymptotically exact in the weak-coupling limit and
equivalent to the leading-order Lang--Firsov approximation in the
strong-coupling regime. For vanishing spin--orbit coupling, the method of
canonical transformations yields a slightly lower ground-state energy than
perturbation theory in the weak-coupling regime. For sufficiently strong
spin--orbit coupling, $V_{s}=t$, the ground-state energy obtained from the
method of canonical transformations lies slightly above the perturbative result.%

\begin{figure}[ptbh]%
\centering
\includegraphics[
height=5.5201in,
width=3.9228in
]%
{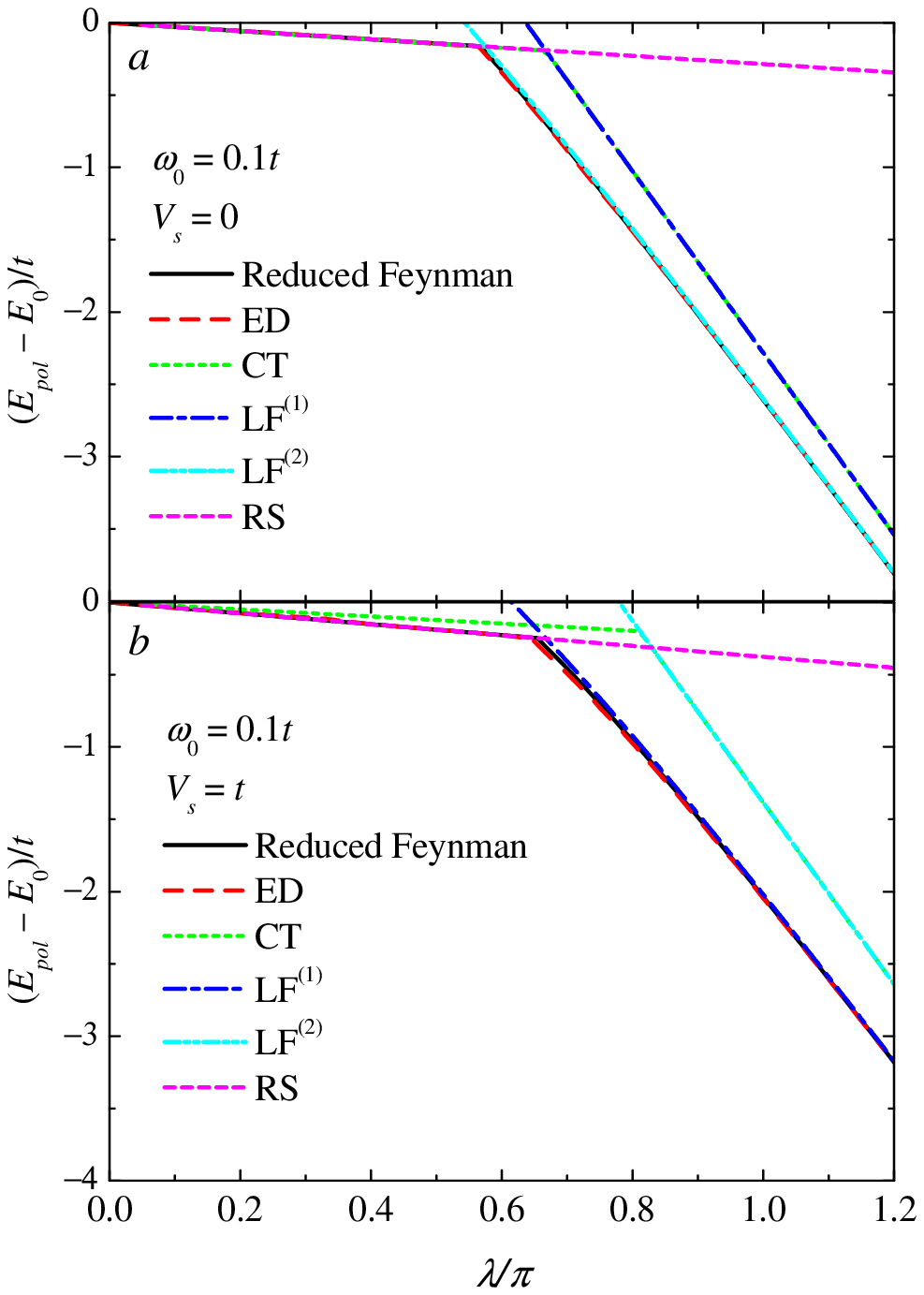}%
\caption{Ground state energy of a polaron with the Rashba SO coupling with the
phonon frequency $\omega_{0}=0.1t$ and the SO potential $V_{s}=0$ (\emph{a})
and $V_{s}=t$ (\emph{b}). The notations are the same as in Fig. \ref{GSSO-1}.}%
\label{GSSO-2}%
\end{figure}

Figure~\ref{GSSO-2} displays the ground-state energy of a polaron with
spin--orbit coupling in the adiabatic regime, $\omega_{0}=0.1t$, calculated
within the reduced Feynman variational approach and compared with the
numerical results of Ref.~\cite{Li2011} and with the same analytic methods as
in Fig.~\ref{GSSO-1}. As can be seen from the figure, in this regime the
reduced Feynman approximation is highly accurate and exhibits good agreement
with the numerical calculation over the entire range of coupling strengths.
This behavior of the variational ground-state energy can be understood from
the fact that the crossover region between the weak- and strong-coupling
regimes narrows as the adiabatic ratio $t/\omega_{0}$ increases.%

\begin{figure}[ptbh]%
\centering
\includegraphics[
height=4.1736in,
width=4.8438in
]%
{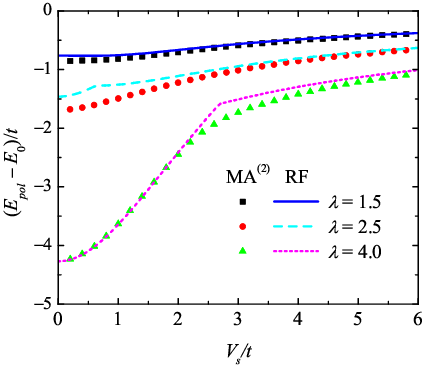}%
\caption{Energy difference $E_{pol}-E_{0}$ as a function of the SO coupling
potential $V_{s}$ for different values of the coupling strength $\lambda$.
\emph{Curves}: results of the calculation within the reduced Feynman
approximation. \emph{Symbols}: results of the momentum-average approximation
MA$^{\left(  2\right)  }$ \cite{Covaci2009}.}%
\label{fig:EnDiffSO}%
\end{figure}

Finally, we examine the dependence of the Rashba polaron ground-state energy
on the spin--orbit coupling potential. In Fig.~\ref{fig:EnDiffSO}, the
difference between the polaron ground-state energy obtained using the reduced
Feynman method and the free-electron ground-state energy is compared with the
momentum-average approximation MA$^{\left(  2\right)  }$ of
Ref.~\cite{Covaci2009} with $\omega_{0}=t$. Overall, good agreement between
the present variational results and the Momentum Average data is observed. The
agreement is particularly good in the strong-coupling regime, which is reached
in Fig.~\ref{fig:EnDiffSO} for $V_{s}\lessapprox2t$. In accordance with the
observations of Ref.~\cite{Covaci2009}, we conclude that spin--orbit coupling
effectively reduces the electron--phonon coupling strength, as evidenced by
the crossover from the strong- to the weak-coupling regime upon increasing the
spin--orbit coupling potential. The transition between the strong- and
weak-coupling regimes occurs at $V_{s}\approx2.7$ for $\lambda=4$, shifting to
a lower value $V_{s}\approx0.6$ for $\lambda=2.5$. For the smallest coupling
constant $\lambda=1.5$, the weak-coupling regime persists for all values of
the SO potential.

\subsection{Polaron dispersion \label{NumDisp}}

\subsubsection{Single band \label{NumDispSO}}

Here, we analyze the polaron dispersion of a Holstein polaron in a single
band, calculated using different analytic approximations.%

\begin{figure}[ptbh]%
\centering
\includegraphics[
height=4.2246in,
width=5.3921in
]%
{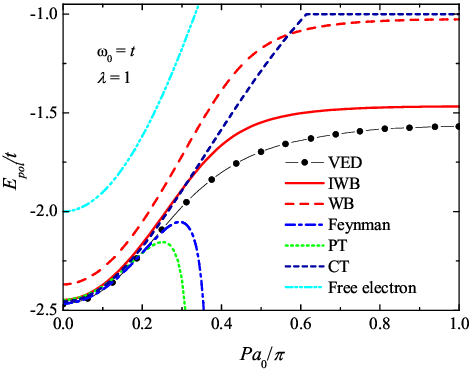}%
\caption{Momentum-dependent polaron self-energy calculated for the Holstein
polaron in 1D for $\omega_{0}=t,$ $a_{0}=1$ and $\lambda=1$ using different
approximations. \emph{Full dots}: Variational Exact Diagonalization (VED)
method. \emph{Dashed red curve}: self-consistent Wigner-Brillouin
approximation (WB). \emph{Solid red curve}: improved WB. \emph{Dot-dashed blue
curve}: modified Feynman variational approach. \emph{Dotted green curve}:
Rayleigh-Schr\"{o}dinger perturbation theory (RS). \emph{Short-dashed blue
curve}: method of canonical transformations (CT). \emph{Dot-dot-dashed cyan
curve}: the free-electron band energy.}%
\label{fig:poldisp1}%
\end{figure}

Figure~\ref{fig:poldisp1} shows the polaron dispersion throughout the
Brillouin zone for the Holstein polaron in 1D with $\omega_{0}=t,$ $a_{0}=1,$
and $\lambda=1,$ obtained using different approximations: the Variational
Exact Diagonalization (VED) method applied by Bon\v{c}a \emph{et al.}
\cite{Bonca1999}, the self-consistent Wigner-Brillouin approximation (WB)
\cite{Berciu2006,Goodvin2006,Goodvin2007,Bhattacharya}, the improved WB as
described in the present work, the modified Feynman variational approach, the
Rayleigh-Schr\"{o}dinger perturbation theory, and the method of canonical transformations.

First, it can be seen from Fig.~\ref{fig:poldisp1} that the
Rayleigh-Schr\"{o}dinger perturbation theory fails to describe the polaron
dispersion over the entire momentum range, being realistic only at small
momenta. The reason is well known: the perturbative expression for the polaron
self-energy at nonzero polaron momentum contains resonances when the momentum
becomes sufficiently large.

A similar feature appears for the momentum-dependent polaron energy calculated
within the modified Feynman variational approach, because the variational
expression contains denominators analogous to those in RS, with the energies
of the trial system replacing the free-electron band energies. Consequently,
the momentum dependence of the polaron energy within RS and within the
modified Feynman variational approach is qualitatively similar, differing
mainly quantitatively.

The self-consistent Wigner-Brillouin expression for the momentum-dependent
polaron self-energy does not contain resonances, because the polaron energy
appears both on the left-hand side and in the denominator on the right-hand
side of the defining equation. The resulting polaron dispersion is therefore
qualitatively realistic over the entire Brillouin zone. However, the polaron
energy within WB is everywhere higher than that obtained within RS, as noted
in many works, in particular in Ref.~\cite{Marchand}.

In order to improve WB while retaining its ability to describe the polaron
dispersion over the full momentum range, Lindemann \cite{Lindemann} and
Gerlach and Smondyrev \cite{Gerlach} proposed a modified WB-like approach that
is variationally correct and reproduces a Tamm-Dankoff-type procedure by
regrouping the unperturbed and interaction Hamiltonians, as described in
Subsec.~\ref{Sec:IWB}. This improved WB matches RS at zero momentum and yields
a polaron dispersion over the entire Brillouin zone. As can be seen from
Fig.~\ref{fig:poldisp1}, the improved WB method gives the polaron self-energy
much closer to the VED results than all other approximations considered in the
present work.

Finally, we also plot in the same figure the polaron self-energy obtained
within the CT approach. While it is rather close to the VED results at small
momentum (performing better than RS and the modified WB), it shows relatively
poor agreement at large momentum. In particular, it exhibits a non-smooth
behavior corresponding to the transition between weak-coupling and
strong-coupling regimes, already indicated above.%

\begin{figure}[ptbh]%
\centering
\includegraphics[
height=4.2255in,
width=5.412in
]%
{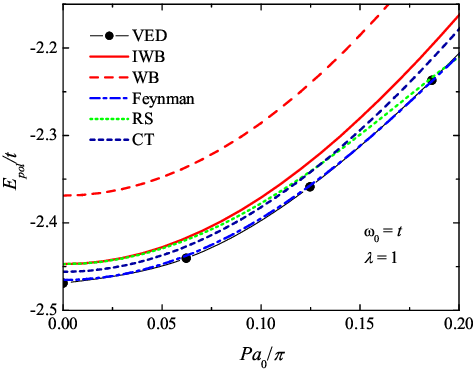}%
\caption{Momentum-dependent polaron self-energy from Fig. \ref{fig:poldisp1}
in the small momentum range. The notations are the same as in Fig.
\ref{fig:poldisp1}.}%
\label{fig:poldisp2}%
\end{figure}

It is also instructive to examine the polaron self-energy in the
small-momentum range, as shown in Fig.~\ref{fig:poldisp2}. As can be seen from
the figure, the non-improved WB polaron energy at small momentum is higher
than that provided by the other approximations. The perturbation theory and
the improved Wigner-Brillouin approximation yield self-energies that tend to
the same limit as $P\rightarrow0$. The polaron self-energy obtained within the
method of canonical transformations at small $P$ is slightly lower than the
perturbative result.

Remarkably, the Feynman variational approach at small momentum is in good
agreement with the Variational Exact Diagonalization method. This is
consistent with the agreement between the Feynman approach and the
Diagrammatic Monte Carlo calculations for the ground-state energy presented in
Subsec.~\ref{NumGS}.%

\begin{figure}[ptbh]%
\centering
\includegraphics[
height=4.0075in,
width=4.7271in
]%
{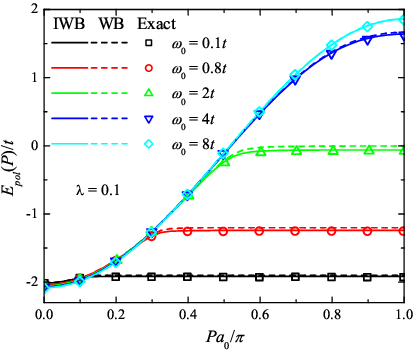}%
\caption{\emph{Symbols}: exact results for the polaron band dispersion of the
1D Holstein model with $\lambda=0.1$ and several phonon frequencies calculated
in Ref. \cite{Wellein1997} for a finite lattice with 20 sites. \emph{Solid and
dashed curves}: results obtained within, respectively, the improved and
standard Wigner-Brillouin approximation.}%
\label{fig:Wellein1}%
\end{figure}

We also refer to the numerically exact calculations of the momentum-dependent
polaron self-energy by Wellein and Fehske \cite{Wellein1997} for comparison
with our results. Figure~\ref{fig:Wellein1} shows the polaron dispersion for
the one-dimensional Holstein model with different phonon frequencies at the
coupling strength $\lambda=0.1$, calculated within the improved (solid curves)
and standard Wigner-Brillouin approximations. The symbols represent data
points from the numerically exact calculations \cite{Wellein1997}.

As can be seen from Fig.~\ref{fig:Wellein1}, the improved Wigner-Brillouin
approximation is in good agreement with the numerically exact results, and
performs systematically better than the standard WB. In addition, a subtle
feature of the dispersion is correctly reproduced: at $P=0$ the polaron
self-energy decreases with increasing $\omega_{0}$, whereas at sufficiently
large momentum the self-energy increases as $\omega_{0}$ increases. This
agreement between analytic and numerical results demonstrates that the WB
procedure removes resonances in a physically correct manner.%

\begin{figure}[ptbh]%
\centering
\includegraphics[
height=4.0075in,
width=4.868in
]%
{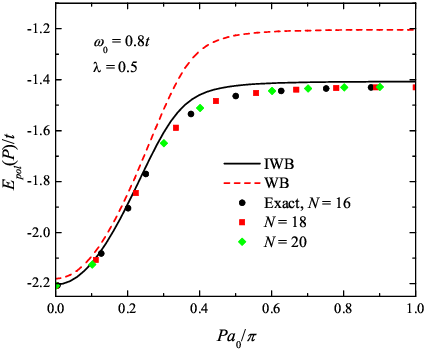}%
\caption{\emph{Symbols}: exact results for the polaron band dispersion of the
1D Holstein model with $\lambda=0.4$ and $\omega_{0}=0.8t$ \cite{Wellein1997}
for a finite lattice with 16, 18 and 20 sites. \emph{Solid and dashed curves}
show results of the present calculation within, respectively, the improved and
standard Wigner-Brillouin approximation.}%
\label{fig:Wellein2}%
\end{figure}

In Fig.~\ref{fig:Wellein2}, a comparison between analytic and numerical
simulations of the momentum-dependent polaron energy is shown for the same
model as in the previous figure, at a stronger coupling $\lambda=0.5$ and
$\omega_{0}=0.8t$. For this coupling strength, a small discrepancy appears
between the exact numerical results and those provided by the improved
Wigner-Brillouin method. The standard WB approximation (dashed curve) performs
poorly at this coupling strength. In summary, the improved WB approximation
appears promising for the description of the polaron dispersion, at least in
the range of weak and intermediate coupling strength.

\section{Conclusions \label{Conclusions}}

In this work, we have developed, extended, and systematically benchmarked a
set of analytic approaches for the polaron problem in a finite-width,
nonparabolic conduction band, with particular emphasis on tight-binding
dispersions and on the interplay between electron--phonon coupling and
spin--orbit interaction. The central motivation of this study was to reassess
the status of well-established continuum polaron methods when the
effective-mass approximation breaks down, and to construct an analytic
framework that remains reliable across weak, intermediate, and strong coupling
regimes without relying on uncontrolled assumptions.

A first important outcome of our analysis is that several approaches
traditionally regarded as \textquotedblleft weak-coupling\textquotedblright%
\ methods exhibit a substantially broader range of validity when reformulated
for a finite-bandwidth lattice polaron. In particular, the self-consistent
Wigner--Brillouin approximation and its improved, variationally consistent
variant provide a momentum-dependent polaron self-energy that is free of
resonance divergences throughout the entire Brillouin zone. The improved
Wigner--Brillouin scheme exactly reproduces Rayleigh--Schr\"{o}dinger
perturbation theory at zero momentum and yields a polaron dispersion in good
agreement with numerically exact results at weak coupling, while remaining
qualitatively reliable up to intermediate coupling strengths. This makes it a
powerful analytic tool for addressing polaron band renormalization, a regime
where bare perturbation theory inevitably fails.

The method of canonical transformations turns out to be particularly revealing
in the lattice context. While its continuum counterpart, the Lee--Low--Pines
approach, is strictly limited to weak coupling, its extension to a
tight-binding band captures both weak- and strong-coupling limits. In the
strong-coupling regime, the method becomes exactly equivalent to the
leading-order Lang--Firsov approximation, thereby providing a single analytic
construction that interpolates between qualitatively distinct physical
regimes. It is known that canonical transformation applied to the Holstein
model gives exact results in the limits of weak and strong coupling, but it
does not capture well the transition region between these two regimes, see for
example Ref. \cite{Cheng2008}. The appearance of a non-smooth crossover
associated with competing variational minima reflects an intrinsic limitation
of the CT method, but at the same time offers a clear physical interpretation
of self-trapping phenomena on a lattice and clarifies why such behavior is
absent in the continuum limit.

The central result of the present work is the extension of the Feynman
variational principle to polarons in nonparabolic, finite-width conduction
bands. We have shown that the Feynman approach can be reformulated without
requiring explicit knowledge of the kinetic-energy functional, by working
entirely in momentum space and by treating the trial system exactly. This
resolves a long-standing conceptual obstacle that had previously restricted
the applicability of the Feynman variational method to parabolic dispersions.
The resulting variational functional yields a rigorous upper bound for the
polaron ground-state energy.

A detailed comparison with numerically exact methods demonstrates that the
modified Feynman variational approach reproduces Diagrammatic Monte Carlo
results for the ground state energy with high accuracy over the entire range
of coupling strengths. Its advantage over the Momentum Average approximation
becomes particularly pronounced at low phonon frequencies, where MA requires
increasingly elaborate refinements to maintain quantitative accuracy. In
contrast, the present approach remains uniformly reliable and smoothly
approaches the adiabatic limit, analytically reducing to the exact Feynman
solution for a continuum polaron as the bandwidth increases. This uniform
performance across weak, intermediate, and strong coupling, as well as across
adiabatic and nonadiabatic regimes, constitutes one of the main strengths of
the method.

It should be noted that, unlike the approximations used here, MA yields the
full spectral weight, thereby providing access to quasiparticle weights and
the spectrum of excited states, and, moreover, to a wide variety of observable
quantities. The full Feynman variational approximation is also a promising
framework for investigating the polaron response, similarly to how this has
been done in large-polaron theory; however, such a study lies beyond the scope
of the present work.

For polarons with Rashba spin--orbit coupling, we have shown that the reduced
Feynman variational approach remains rigorously justified and yields
ground-state energies in good agreement with numerically exact calculations.
The method correctly captures the effective reduction of the electron--phonon
coupling induced by spin--orbit interaction and reproduces the continuous
crossover between weak and strong coupling as functions of both the
electron--phonon and spin--orbit coupling strengths. Comparison with
Diagrammatic Monte Carlo data confirms the quantitative reliability of the
approach, particularly in the adiabatic and strong-coupling regimes, where
competing analytic methods become less controlled. The extension of the
analytical framework to polarons with Rashba-type spin-orbit coupling provides
a stringent test of the methods in the presence of nontrivial band topology
and complex matrix Hamiltonians, confirming their robustness beyond
single-band models.

Beyond its quantitative success, the present formulation of the Feynman
variational method has a broader methodological significance. It provides a
consistent and flexible framework for treating polarons in arbitrary lattice
bands, independent of the specific form of the electron--phonon interaction or
phonon dispersion. The approach can be straightforwardly extended to multiband
systems, and to more general forms of spin--orbit coupling. It also opens the
way to controlled treatments of nonlinear electron--phonon interactions.

In summary, the results presented here demonstrate that the Feynman
variational principle, when properly reformulated, remains one of the most
powerful and versatile analytic tools for the polaron problem. Its extension
to nonparabolic, finite-width conduction bands closes an important conceptual
gap between continuum and lattice polaron theories and provides a unified
analytic description that is competitive with state-of-the-art numerical
methods across a wide range of physically relevant regimes.

\begin{acknowledgments}
We are grateful to Mona Berciu for Diagrammatic Monte Carlo and
momentum-average data used for the comparison with the analytic predictions of
the present work. S.K., M.H., I.Z. and J.T. acknowledge the Research
Foundation Flanders (FWO), file numbers 1224724N, 1120825N and V472923N, for
their funding of this research. C.F., J.T., S.R. and T.H. acknowledge support
from the joint Austrian Science Fund (FWF) - FWO project 10.55776/PIN5456724.
A.S.M. and S.R. acknowledge support from the Croatian Science Foundation
(HRZZ) under the project No. IP-2024-05-2406. L.C. acknowledges the Vienna
Doctoral School of Physics. Our Diagrammatic Monte Carlo computational results
presented in this article were obtained using the Austria Scientific Cluster (ASC).
\end{acknowledgments}

\section*{Data availability statement}

The numerical data supporting the findings of this work are publicly available
in \cite{Numdata}.

\end{document}